\documentclass[twocolumn]{aa}
\usepackage{amsmath}
\usepackage{natbib}
\usepackage{graphicx}
\usepackage{txfonts}
\usepackage{subfigure}
\usepackage{pstricks,pst-node}
\usepackage{pst-pdf}

\bibpunct{(}{)}{;}{a}{}{,}

\begin{document}

\title{Bayesian analysis of polarization measurements}
\author{Jason L. Quinn}

\institute{Pontificia Universidad Cat\'{o}lica de Chile\\
           Departamento de Astronom\'{i}a y Astrof\'{i}sica\\
           Vicu\~{n}a Mackenna 4860\\
           Macul 690-4411\\
           Santiago, Chile \\
           \email{jquinn@astro.puc.cl}
          }

\date{Received XXX XX, 20XX; accepted XXX XX, 20XX}

\abstract
{A detailed and formal account of polarization measurements using Bayesian analysis
is given based on the assumption of gaussian error for the Stokes parameters. This
analysis is crucial for the measurement of the polarization degree and angle
at low and high signal-to-noise. The treatment serves as a framework for customized
analysis of data based on a particular prior suited to the experiment.}
{The aim is to provide a rigorous and self-consistent Bayesian treatment of polarization measurements and their statistical error focused
on the case of a single measurement.}
{Bayes Theorem is used to derive a variety of posterior distributions for polarization measurements.}
{A framework that may be used to construct accurate polarization point estimates and confidence
intervals based on Bayesian ideas is given. The results may be customized for a prior and loss
function chosen for a particular experiment.
}
{}


\maketitle

\section{Introduction}
Measuring polarization presents several challenges.
The most important are that a naive calculation
of small polarization seems to be biased towards higher values \citep{1958AcA.....8..135S}, the error bars of
the measured polarization are non-symmetric below
a threshold signal-to-noise \citep{1985A&A...142..100S}, and the
angular distribution is non-gaussian \citep{1993A&A...274..968N}. Given measured values of
the Stokes parameters, what are the best estimates for the true values of
the polarization degree and angle and how
should error bars be assigned?

The polarization state of quasi-monochromatic light may be stated
in terms of $I_0$, $Q_0$, $U_0$, and $V_0$ Stokes parameters which
are defined as certain averages of the electric field along
a pair of orthogonal axes perpendicular to the direction of
the light's propagation (see, e.g., \cite{2003isp..book.....D} or \cite{2002apsp.conf....1L}
for more information). $I_0$ is the intensity, $Q_0$ and $U_0$ are related to linear
polarization, and $V_0$ is circular polarization. The Stokes parameters
are subject to two conditions:
\begin{equation}
I_0^2 \ge Q_0^2 + U_0^2 + V_0^2
\label{condition1}
\end{equation}
and
\begin{equation}
I_0 \ge 0.
\end{equation}
If incoming light is totally polarized, then the equality holds in Eq.~\ref{condition1},
otherwise it is partially polarized or unpolarized. To keep the
equations of manageable size, the rest of this paper
makes the assumption that circular polarization is negligible (i.e., $V_0=0$). It should be straight-forward
to generalize the results to include a non-zero $V_0$.
This restriction is not very severe because circular polarization is
small compared to linear polarization for many astrophysical sources.
It will also be taken that the Stokes parameters are constant
during the course of a single observation.

Stokes parameters are calculated from intensities that follow a Poisson
distribution governed by photon counts. \citet{1983A&A...126..260C} and
\citet{1992A&A...260..525M} have investigated this to make optimal
estimates of them from repeated measurements. In general, however,
a full Poisson treatment is computationally difficult when large
intensities are involved. To avoid concerns about Poisson statistics,
it is assumed that all intensities used to calculate $Q$ and $U$
(and also $I$) are large enough to be treated as having gaussian distributions.

The specifics regarding the calculation of $I$, $Q$, $U$ and their
errors ($\Sigma{}_I$, $\Sigma{}_Q$, $\Sigma{}_U$) depend on the experimental design \citep{2002apsp.conf..303K}.
For example, if one is using a dual-beam polarimeter
with a half-wave-plate, \cite{2006PASP..118..146P} present an optimal set of equations for
calculating the Stokes parameters from the intensities of the ordinary
and extraordinary beams at different waveplate angles. Since the assumption
that these intensities are large enough to be approximated by a Gaussian is being
made, this implies that one may use the usual Gaussian error propagation
formula for $\Sigma{}_I$, $\Sigma{}_Q$, and $\Sigma{}_U$ based on the errors
for the individual intensities (square root of the counts) that went into their
calculation. Regardless how $I$, $Q$, and $U$ are actually calculated, they may be treated
as logically separate (e.g., one could imagine one instrument measuring
intensity with error $\Sigma{}_I$ in conjunction with completely separate instruments measuring
$Q$ and $U$ with errors $\Sigma{}_Q$ and $\Sigma{}_U$, respectively). Hereon,
the error on the measured total intensity, $\Sigma{}_I$, is treated as being
independent of $\Sigma{}_Q$ and $\Sigma{}_U$.

Most of the polarization error literature makes
the further assumption that the measured Stokes parameters are normally
distributed about their true values, $Q_0$ and $U_0$, with standard
deviation $\Sigma{}_Q$ and $\Sigma{}_U$. It is also usually assumed that
the $Q$ and $U$ distributions are characterized by equal dispersion. Unfortunately, $\Sigma{}_Q=\Sigma{}_U$
is not generally satisfied for arbitrary datasets. This assumption is a pragmatic
one to avoid ellipsoidal distributions which
would complicate the analysis. Careful design of the experiment can make it
such that $\Sigma{}_Q \approx \Sigma{}_U$. Extension of the results to allow
for unequal variance is reserved for future work. It is now
assumed that $\Sigma{}_Q$ and $\Sigma{}_U$ are equal and called $\Sigma{}$.
The quantity $\Sigma{}_I$ only tangentially enters the analysis, leaving
$\Sigma{}$ as the lone important error quantity. In practice, $\Sigma{}$
could be set equal to the average of $\Sigma{}_Q$ and $\Sigma{}_U$ or
the maximum value if one is more conservative.

The units on $I$, $Q$, $U$ and $\Sigma{}_I$, $\Sigma{}_Q$, $\Sigma{}_U$
are energy per time. Shortly, new variables will be defined that are unitless.
They will be best interpreted as percentages or ratios and the issue of units vanishes.

\section{Theoretical background}

In this section, the various distributions that will be needed are derived.
The distributions will be presented in several different coordinate
systems because this is intended to serve as a reference and
because it is exceedingly easy to miss factors introduced by the Jacobian.

\subsection{The Sampling Distribution}
\subsubsection{Variable introductions and large intensity limit}
The fundamental assumption of this paper is that the measurement of
the (unnormalized) Stokes parameters, $Q$ and $U$, which are assumed
to be uncorrelated, is described by a two-dimensional gaussian
with equal standard deviation, $\Sigma{}$, in both directions.
This distribution, $F_C$, is
\begin{equation}
F_C(Q,U|Q_0,U_0,\Sigma{}) = \frac{1}{2\pi{}\Sigma{}^2} \operatorname{exp}\left(-\frac{(Q-Q_0)^2+(U-U_0)^2}{2\Sigma{}^2}\right).
\label{eq:unnormalized}
\end{equation}
(Most of the work for this paper will be done in polar coordinates. A $C$-subscript
will be used for distributions in Cartesian coordinates.)
The value of $\Sigma$ is a positive constant assumed to be known precisely and, in practice,
is estimated from the data. The parameters, $Q_0$ and $U_0$, must be elements of a disk of radius $I_0$ centered on
the origin in the $Q_0$-$U_0$ plane. The range of both $Q$
and $U$ is $(-\infty,\infty)$ due to measurement error although the probability
of measuring a value outside a disk of radius $I_0$ diminishes rapidly even for
values of $Q_0$ and $U_0$ near the rim. Eq.~\ref{eq:unnormalized}
is normalized ($  \int_{-\infty}^{\infty} \! \int _{-\infty}^{\infty} F_C \, dQ dU = 1$).

It is often helpful to work in normalized Stokes parameters.
Define new variables $q \equiv Q/I_0$, $u \equiv U/I_0$, $q_0 \equiv Q_0/I_0$, $u_0 \equiv U_0/I_0$,
and $\sigma \equiv \Sigma{}/I_0$. 
The scaled error, $\sigma$, is an admixture between the intensity and the mean $Q$-$U$ error.
It or its inverse, $1/\sigma{}$ =$I_0/\Sigma{}$, may be regarded as a measure of data quality
(cf. $I_0/\Sigma{}_I$, $Q/\Sigma{}_Q$, and $U/\Sigma{}_U$). Small $\sigma$ (large $1/\sigma$) implies
good data. The values for $q_0$ and $u_0$
are restricted to a unit disk centered on the origin. The new probability
density, $f_C$, after the change of coordinates is
\begin{equation}
f_C(q,u|q_0,u_0,\sigma{}) = \frac{1}{2\pi{}\sigma{}^2} \operatorname{exp}\left(-\frac{(q - q_0)^2+(u - u_0)^2}{2 \sigma{}^2}\right).
\label{msimple}
\end{equation}
This equation is also normalized ($  \int_{-\infty}^{\infty} \! \int _{-\infty}^{\infty} f_C \, dq du = 1$).

The normalized Stokes parameters $q$ and $u$ present a minor problem. How
can they be calculated if $I_0$ is an unknown quantity? It might be thought
that the definition of $q$ and $u$ should use the measured intensity, $I$, instead of the
true intensity, $I_0$. If that alternative
definition is used, however, the differentials of $q$ and $u$
are more complicated than just $dq = dQ/I_0$ and $du = dU/I_0$. This makes a change
of variables for the distribution difficult. It is better to keep
the first definition and require that $I_0 \approx I$. This occurs
when the total number of counts that went into the measurement of
$I$ is large, which also implies $\Sigma_I \ll I$.

Another helpful set of variables are ``signal-to-noise'' ratios. They are defined by
$\overline{q} \equiv q/\sigma{}$, $\overline{u} \equiv u/\sigma{}$, $\overline{q}_0 \equiv q_0/\sigma{}$, and $\overline{u}_0 \equiv u_0/\sigma{}$.
In these barred variables, the new distribution, $\overline{f}_C$, is
\begin{equation}
\overline{f}_C(\overline{q},\overline{u}|\overline{q}_0,\overline{u}_0) = \frac{1}{2\pi{}} \operatorname{exp}\left(-\frac{(\overline{q} - \overline{q}_0)^2+(\overline{u} - \overline{u}_0)^2}{2}\right).
\label{eq:normalizedsnr}
\end{equation}
This time, the Jacobian causes the $1/\sigma^2$ leftover after the change
of variables to disappear. This is normalized as well
($  \int_{-\infty}^{\infty} \! \int _{-\infty}^{\infty} \overline{f}_C \, d\overline{q} d\overline{u} = 1$);
but, in this case, $\overline{q}_0$ and $\overline{u}_0$ are restricted to a disk of radius $1/\sigma{}$.

\subsubsection{Transformation to polar coordinates}
The true polarization degree, $p_0$, and true angle on the sky, $\phi{}_0$, may be defined in terms of
the true (i.e., perfectly known) $q_0$ and $u_0$ Stokes parameters by
\begin{equation}
p_0 \equiv \sqrt{q_0^2 + u_0^2}   \;\;\;\; \text{and}  \;\;\;\;   \phi{}_0 \equiv \frac{1}{2} \operatorname{atan}{\left(\frac{u_0}{q_0}\right)}.
\label{eq:polartrans}
\end{equation}
The true angle on the sky is related to the true angle in the $q_0$-$u_0$ plane by the relation $\theta{}_0 \equiv 2\phi{}_0$.
Similarly,
\begin{equation}
p \equiv \sqrt{q^2 + u^2}   \;\;\;\; \text{and}  \;\;\;\;   \phi{} \equiv \frac{1}{2} \operatorname{atan}{\left(\frac{u}{q}\right)}
\label{eq:nosubpolartrans}
\end{equation}
and $\theta{} \equiv 2\phi{}$. The range of $p$ is $[0,\infty)$ and the range of $p_0$ is $[0,1]$. Define
$\overline{p} \equiv p/\sigma{}$ and $\overline{p}_0 \equiv p_0/\sigma{}$. It is important to notice that the
range of $\overline{p}_0$ is now $[0,1/\sigma{}]$ or $[0,I_0/\Sigma{}]$ because
in the Bayesian analysis, this will be the range of integration.
The polarization angle defines the plane of vibration of
the electric field on the sky. By convention, it is usually
defined such that $0^\circ{}$ corresponds to North and
increases eastward such that East is $\phi{}=90^\circ{}$.

Eq.~\ref{msimple} may be converted to polar coordinates by inverting
Eqs.~\ref{eq:polartrans} and \ref{eq:nosubpolartrans}. Using $\theta{}$
instead of $\phi{}$, this yields $q=p \cos{\theta{}}$ and $u=p \sin{\theta{}}$ and
corresponding equations for the zero-subscripted true values. The normalized probability density distribution in polar coordinates $(p,\theta{})$ is
\begin{equation}
f'(p,\theta{}|p_0,\theta{}_0,\sigma) = \frac{p}{2\pi{}\sigma{}^2} \operatorname{exp}\left(-\frac{p^2+p_0^2-2pp_0\cos{(\theta{}-\theta{}_0)}}{2 \sigma{}^2}\right).
\label{eq:polar}
\end{equation}
A prime is used as part of the function's name to serve as a warning that
the angle in the $q$-$u$ plane, $\theta$, is being used as the angular variable.
The factor of $p$ in front enters because $dq \, du = p \, dp \, d\theta{}$ (i.e., due to the Jacobian
of the transformation). Some authors choose to leave the factor of
$p$ off of the  $p,\theta{}$-distribution and insert it manually when needed.
Probability distributions transform as a scalar density under
coordinate transformations so they should ``pick-up'' the Jacobian factor. Caution is needed
when reading the literature.

\begin{figure}
\centering
\subfigure[$p_0/\sigma{}=0.0$ and $\phi{}_0=0$.]
{
    \label{fig:f_plot_a}
    \includegraphics[width=8.75cm]{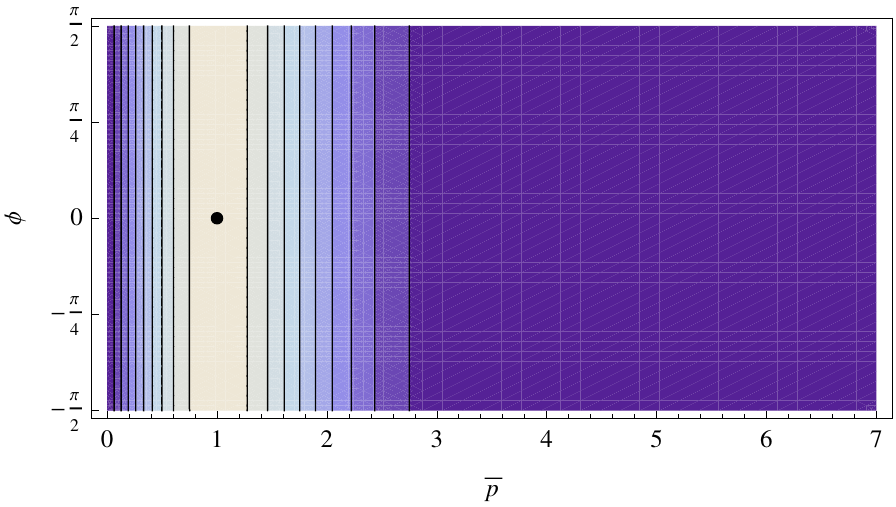}
}
\hspace{1cm}
\subfigure[$p_0/\sigma{}=0.5$ and $\phi{}_0=0$.]
{
    \label{fig:f_plot_b}
    \includegraphics[width=8.75cm]{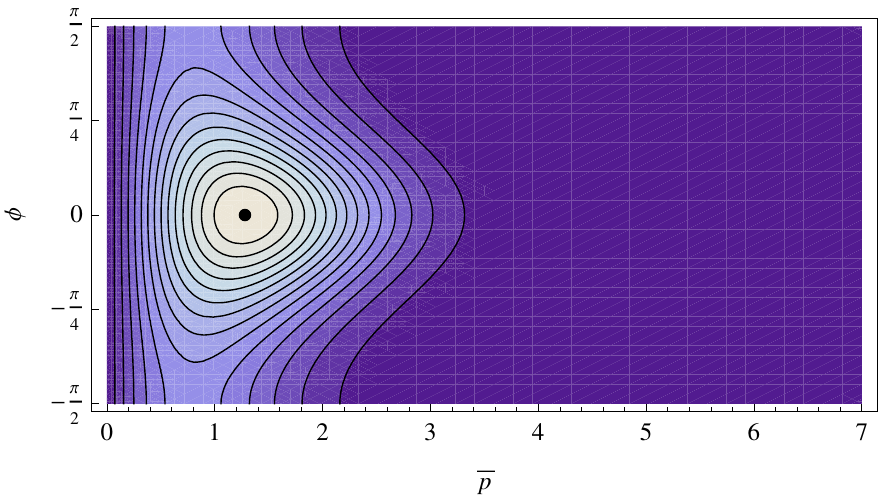}
}
\hspace{1cm}
\subfigure[$p_0/\sigma{}=2.0$ and $\phi{}_0=0$.]
{
    \label{fig:f_plot_c}
    \includegraphics[width=8.75cm]{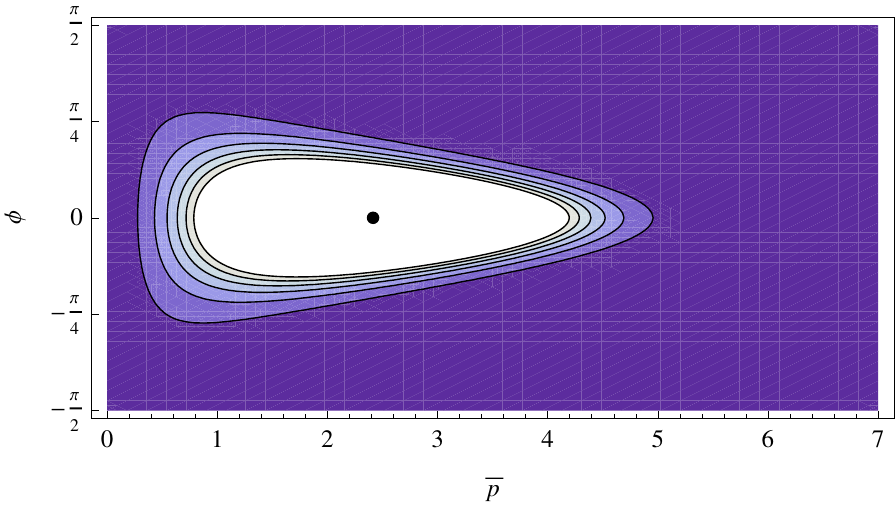}
}
\caption{Contour plots of $\overline{f}(\overline{p},\phi{}|\overline{p}_0,\phi{}_0)$ (Eq.~\ref{eq:polarSNR}) for a few different values of $p_0/\sigma{}$ with $\phi{}_0=0$.
The dots indicate the maximums of the distributions. Contour spacings are at intervals of $0.02$.}
\label{fig:angleplot}
\end{figure}

The distribution in terms of the sky angle is
\begin{equation}
f(p,\phi{}|p_0,\phi{}_0,\sigma) = \frac{p}{\pi{}\sigma{}^2} \operatorname{exp}\left(-\frac{p^2+p_0^2-2pp_0\cos{(2(\phi{}-\phi{}_0))}}{2 \sigma{}^2}\right)
\label{eq:polar2}
\end{equation}
and in barred variables,
\begin{equation}
\overline{f}(\overline{p},\phi{}|\overline{p}_0,\phi{}_0) = \frac{\overline{p}}{\pi{}} \operatorname{exp}\left(-\frac{\overline{p}^2+\overline{p}_0^2-2\overline{p} \, \overline{p}_0\cos{(2(\phi{}-\phi{}_0))}}{2}\right).
\label{eq:polarSNR}
\end{equation}
The $\overline{f}(\overline{p},\phi{}|\overline{p}_0,\phi{}_0)$ distribution is normalized under
the support $[0,\infty) \times{} (-\pi/2,\pi/2]$ (i.e., $ \int_{-\frac{\pi}{2}}^{\frac{\pi}{2}} \! \int _{0}^{\infty} \overline{f} \, d\overline{p} d\phi = 1   $). The origin has
infinitesimal measure so usually it does not contribute finitely to an integral.
In later sections, integration over delta functions (aka delta distributions) centered
at the origin will occur. This requires some formality in the polar coordinate definition.
The parameter space technically has $(p=0,\phi)$ identified for all $\phi$ as the origin and
$(p,\phi{})$ is identified with $(p,\phi=\phi+n\pi{})$ for all $p>0$ and any integer $n$.
The resulting quotient space has $[0,\infty) \times{} (-\frac{\pi}{2},\frac{\pi}{2}]$ as a fundamental cell.
Functions on this space are periodic in $\phi$ and they must satisfy $f(p,\phi)=f\left(p,((\phi+\frac{\pi}{2}) \mod \pi) -\frac{\pi}{2}\right)$.
Here, ``mod'' is the modulus operator. The $\pi/2$ operations perform shifts needed
because the angles are defined as $(-\frac{\pi}{2},\frac{\pi}{2}]$ instead of $[0,\pi)$. This formality
is sometimes important in calculations, like median estimates of $\phi_0$, where
it is natural to have the range of integration extend outside of the
$(-\frac{\pi}{2},\frac{\pi}{2}]$ domain.

Some intuition is gained by examining the distribution in Eq.~\ref{eq:polarSNR}
for differing values of $p_0/\sigma{}$. Figs.~\ref{fig:f_plot_a}, \ref{fig:f_plot_b}, and \ref{fig:f_plot_c} show
the distributions for $\phi{}_0=0$ and $p_0/\sigma{}=0.0$, $0.5$ and $2.0$, respectively. Choosing a different value of
$\phi{}_0$, for a given value of $p_0/\sigma{}$, only translates
the distribution in the $\phi$-direction and does not change its shape; therefore, all plots simply use $\phi{}_0=0$.
When $p_0/\sigma{}=0$ the angular distribution is flat and much of the probability is concentrated
in a band around $p/\sigma{}=1$. As $p_0/\sigma{}$ increases, an oval-shaped ``probability bubble'' forms. This shape persists even at large values. The probability under the distribution to the left and right of the maximum is asymmetric, with more
to the left. As $p_0/\sigma{}$ continues to increase, the probability
on each side of the maximum approaches $0.5$. This is true even though the oval shape stays present.

The sampling distribution for $N$ measurements is simply
a product of those for individual measurements when
the measurements are independent of one another,
\begin{equation}
\mathcal{F}(\{p,\phi{}\}_N|p_0,\phi{}_0,\sigma) = \displaystyle\prod_{n=1}^{N} f((p,\phi{})_n|p_0,\phi{}_0,\sigma).
\end{equation}
In this notation, $(p,\phi{})_n$ denotes the $n$-th measurement and $\{p,\phi{}\}_N$ is
the whole set of measurements. This paper
focuses on the case of a single measurement, such that $\mathcal{F}=f$.

\subsection{The ``most probable'' and maximum likelihood estimators}
Two classical analytic estimators for the unknown parameters are easily found from
the sampling distribution. These are the maximum likelihood and
``most probable'' estimators. Later new estimators will be introduced
using the posterior distribution.

If $f(p,\phi{}|p_0,\phi{}_0,\sigma)$ is viewed as
a function of $p_0$ and $\phi{}_0$ for a fixed $p$ and $\phi$, it is called the 
likelihood function. Solving the system $\frac{\partial{f}}{\partial{p_0}}=0$ and $\frac{\partial{f}}{\partial{\phi_0}}=0$ yields
the maximum likelihood estimate (ML),
\begin{equation}
\hat{p}_{0,ML} = p \quad  \text{and} \quad \hat{\phi}_{0,ML} = \phi{}.
\label{eq:ML}
\end{equation}
This solution is not particularly useful as it does not correct
for bias. There is potential for confusion here: if
the same construction is applied to the sampling
distribution marginalized over the angle (i.e., Rice
distribution), the maximum likelihood
estimator \textit{does} correct for some bias. The
Rice distribution will be covered in Section~\ref{sec:rice}.

\citet{1997ApJ...476L..27W} suggest simultaneously
maximizing Eq.~\ref{eq:polar} with respect to $p$ and $\theta{}$ to
estimate $p_0$ and $\theta{}_0$. This yields another estimator
sometimes called the ``most probable'' estimator (MP)
in the literature (or Wardle and Kronberg estimator when applied
to the Rice distribution). The system
$\frac{\partial{f}}{\partial{p}}=0$ and $\frac{\partial{f}}{\partial{\phi}}=0$
has the following solution for a maximum:
\begin{equation}
\hat{p}_{0,MP}=\left(p-\frac{\sigma^2}{p}\right) h(p-\sigma) \quad \text{and} \quad \hat{\phi}_{0,MP}=\phi,
\label{eq:wang}
\end{equation}
where $h(x)$ is the Heaviside step function. The Heaviside
function can be motivated by contemplation of Fig.~\ref{fig:f_plot_a}.
Eq.~\ref{eq:wang} may be used to find the maximums in Fig.~\ref{fig:angleplot}.
It is worth noting that this formula is also expressible as
$\hat{p}_{0,MP} = \left(\left( p - \frac{\sigma{}^2}{p} \right)    +     \left| p - \frac{\sigma{}^2}{p} \right|\right)/2$,
which may be more practical in data reduction pipelines,
and $\hat{\overline{p}}_{0,MP} = \left(\overline{p}-\frac{1}{\overline{p}}\right)h(\overline{p}-1)$
in terms of the barred variables.

The goal is to replace these estimators
with new ones based on Bayesian ideas and Decision
Theory. The ML estimate continues to be relevant in that context
because it is equal to some Bayesian estimators when a uniform prior is
used.

\subsection{The Rice distribution}
\label{sec:rice}
The Rice distribution \citep{1945BSTJ...24...46R}, $R$, is the marginal distribution
of $p$ over $f$,
\begin{equation}
R(p|p_0,\sigma{}) = \int_{-\pi{}/2}^{\pi{}/2} \! f(p,\phi{}|p_0,\phi{}_0,\sigma)  \,  d\phi{}.
\end{equation}
This integration may be accomplished using the
integral $\int_{-\pi/2}^{\pi/2} \! e^{a \cos{(2x)}} \, dx = \pi{} \mathcal{I}_0(a)$ so that
\begin{equation}
R(p|p_0,\sigma{}) = \frac{p}{\sigma{}^2} \operatorname{exp}\left(-\frac{p^2+p_0^2}{2\sigma{}^2}\right)   \mathcal{I}_0\left(\frac{pp_0}{\sigma{}^2}\right).
\label{eq:rice}
\end{equation}
Here, $\mathcal{I}_0(x)$ denotes the zeroth-order modified Bessel function
of the first kind.\footnote{Some papers in the literature use
Bessel functions, $\mathcal{J}$, instead of \textit{modified} Bessel functions,
$\mathcal{I}$. This introduces distracting complex factors of $i$ into the equations. It is
worth noting that $\mathcal{J}_0(i x) = \mathcal{I}_0(x)$ and $\mathcal{J}_1(i x) = i \mathcal{I}_1(x)$ to
aid in reading those papers.} It is an even \textit{real} function when its argument is real
and $\mathcal{I}_0(0)=1$. The Rice distribution does not depend on $\phi{}_0$ or $\phi{}$ after
the integration over $\phi{}$ so they have been dropped from the notation.

\begin{figure}
\centering
\subfigure[A view of the Rice distribution near the origin. The value of $\sigma$ is mostly irrelevant
so long as $1/\sigma$ is at least $3$ such that the domain plotted is valid. For instance, it may
be taken to be $100$ for comparison with later figures. The meaning of the
dashed and dotted curves and the dots is discussed in the text.]
{
    \label{fig:contour_rice}
    \includegraphics[width=8.75cm]{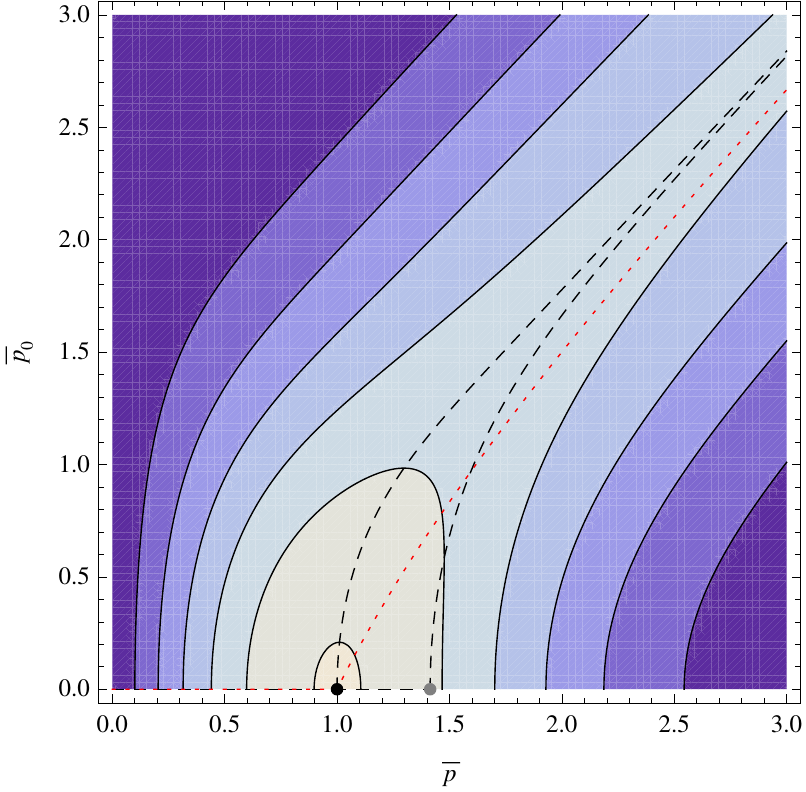}
}
\hspace{1cm}
\subfigure[A view of the Rice distribution for the full domain of $\overline{p}_0$ when $\sigma=1/8$.
Notice that the linear-like band that develops persists to the maximum value of
$\overline{p}_0$, which is $8$ in this case. This is different than the behavior
of the posterior distribution as will be seen.]
{
    \label{fig:contour_rice_panel_two}
    \includegraphics[width=8.75cm]{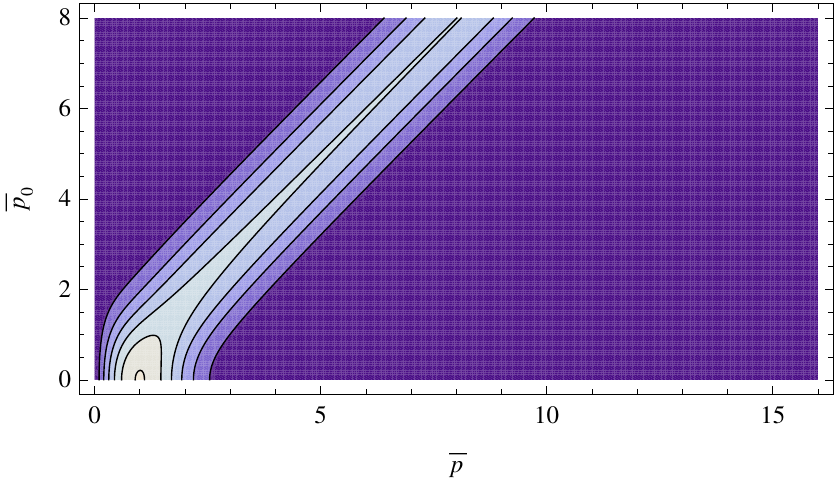}
}
\caption{Two contour plots of the Rice distribution (Eq.~\ref{eq:riceSNR}). The contours are spaced by
0.1 intervals. The value of $\sigma$ is mostly unimportant for $\overline{R}$. It does
not affect the shape of the distribution. The only purpose of mentioning it is to insure
$\overline{p}$ exists for values above $1/\sigma$.}
\label{fig:contourR}
\end{figure}

In barred variables, the Rice distribution is
\begin{equation}
\overline{R}(\overline{p}|\overline{p}_0) = \overline{p} \operatorname{exp}\left(-\frac{\overline{p}^2+\overline{p}_0^2}{2}\right)   \mathcal{I}_0\left(\overline{p} \, \overline{p}_0\right).
\label{eq:riceSNR}
\end{equation}
This is shown as a contour plot in Fig.~\ref{fig:contour_rice}. In this plot, $\overline{R}$
is plotted against $\overline{p}_0$ and $\overline{p}$. The contours are at $0.1$ intervals. There are
two special points in the plot. A global maximum of $1/\sqrt{e}\approx{}0.6065$ is reached at
$(1,0)$ (black dot) and a critical point at $(\sqrt{2},0)$ (gray dot) is found if maximums along vertical slices are examined.
Recall $\overline{p}=p/\sigma{}$ and $\overline{p}_0=p_0/\sigma{}$.
The maximum along a vertical slice in the interval $0\le{}p/\sigma{}\le{}\sqrt{2}$ lies on the
$p/\sigma{}$-axis while for values greater than $\sqrt{2}$ it lies above the axis. Tracing the maximum of the vertical slices for $\frac{p}{\sigma{}}>\sqrt{2}$ implicitly defines a curve
which is given by
\begin{equation}
\overline{p} \mathcal{I}_1(\overline{p} \, \overline{p}_0) - \overline{p}_0 \mathcal{I}_0(\overline{p} \, \overline{p}_0)=0.
\label{eq:rice_vertical_slice}
\end{equation}
This is the ML estimator of \cite{1985A&A...142..100S} (with a factor of $i$ corrected).
$\mathcal{I}_1(x)$ is the first-order modified Bessel function of the first kind. It is an odd real function
when its argument is real.
Similarly, horizontal slices define a curve via
\begin{equation}
(\overline{p}^2-1)\mathcal{I}_0(\overline{p} \, \overline{p}_0) - \overline{p} \, \overline{p}_0 \mathcal{I}_1(\overline{p} \, \overline{p}_0)=0.
\label{eq:rice_horizontal_slice}
\end{equation}
This is the Wardle and Kronberg estimator \citep{1974ApJ...194..249W}.
These are the two dashed curves in the plot. The curve for the horizontal slice
maximums is the one that terminates at the global maximum. The red dotted line
is the Wang estimator from Eq.~\ref{eq:wang}. There is a significant difference
between the Rice distribution-based estimators and the two-dimensional estimator.
For large values of $\overline{p}$, it estimates a smaller value of $\overline{p}_0$ lower than
the two other curves.

Fig.~\ref{fig:contour_rice_panel_two} plots the Rice distribution with the full range
of $\overline{p}_0$ when $\sigma=1/8$. Except near the origin, much of the probability is distributed in a diagonal
linear band that continues uninterrupted until the maximum value of $\overline{p}_0$. Later
it will be seen that the posterior distribution becomes non-linear for large values of
polarization.

\subsection{The Posterior Distribution}
The distribution given by Eq.~\ref{eq:polar2} is the distribution
of the measured values given true values as input parameters. In practice,
one usually wishes to estimate the true values from the measured values.
This is accomplished through the posterior distribution, $B$, given by Bayes Theorem provided one accepts
some prior distribution, $\kappa{}(p_0,\phi{}_0)$, for the model parameters. The posterior
distribution (for a single measurement) is,
\begin{equation}
B(p_0,\phi{}_0|p,\phi{},\sigma) = \frac {f(p,\phi{}|p_0,\phi{}_0,\sigma{}) \kappa{}(p_0,\phi{}_0)}
                  { \int_{-\frac{\pi}{2}}^{\frac{\pi}{2}} \! \int_{0}^{1} \! f(p,\phi{}|p'_0,\phi{}'_0,\sigma{})   \kappa{}(p'_0,\phi{}'_0)      \,  dp'_0  \,  d\phi{}'_0 }.
\label{eq:posterior}
\end{equation}

Care must be taken with the limits of the $p_0$ integration and it
must always be remembered if one is using the unnormalized, normalized,
or barred variables. The barred version is
\begin{equation}
\overline{B}(\overline{p}_0,\phi{}_0|\overline{p},\phi{},\sigma) = \frac {\overline{f}(\overline{p},\phi{}|\overline{p}_0,\phi{}_0) \overline{\kappa{}}(\overline{p}_0,\phi{}_0)}
                  { \int_{-\frac{\pi}{2}}^{\frac{\pi}{2}} \! \int_{0}^{1/\sigma{}} \overline{f}(\overline{p},\phi{}|\overline{p}'_0,\phi{}'_0)   \overline{\kappa{}}(\overline{p}'_0,\phi{}'_0)      \,  d\overline{p}'_0  \,  d\phi{}'_0 }
\label{eq:posteriorSNR}
\end{equation}
with $\overline{\kappa}(\overline{p}_0,\phi_0)=\sigma \kappa(\sigma{} \overline{p}_0,\phi{})$. It is critical to notice the upper limit of the integration
of $\overline{p}_0$ is $1/\sigma{}$, which equals $I_0/\Sigma{}$. It is not infinity
as is sometimes used. In practice, however, many of the integrals over $\overline{p}_0$ that
occur in Bayesian polarization equations, tend to be extremely insensitive to the
value of $1/\sigma{}$ so long as it is above a value of about $4$. Thus, using
infinity as the upper limit often produces a very reasonable approximation.
This has the added advantage that integrals sometimes have explicit closed forms
when otherwise they might not.

There is great freedom in defining a Bayesian prior. The function is ostensibly
required to be a probability distribution.
This necessitates that $\kappa(p_0,\phi_0)$ be non-negative everywhere;
however, in most situations, the additional requirement that the distribution be
normalizable may be relaxed as the normalization constant cancels. The choice
of prior will be further discussed in Section~\ref{Elicitation}.

\begin{figure}
\centering
\subfigure[$p/\sigma{}=0.0$ and $\phi{}=0$.]
{
    \label{fig:Bj_plot_a}
    \includegraphics[width=8.75cm]{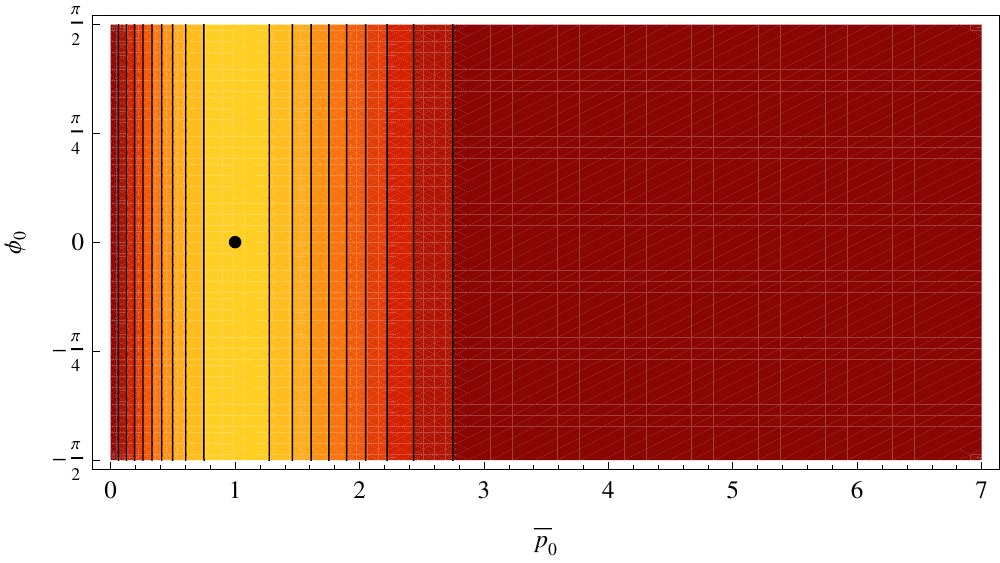}
}
\hspace{1cm}
\subfigure[$p/\sigma{}=0.5$ and $\phi{}=0$.]
{
    \label{fig:Bj_plot_b}
    \includegraphics[width=8.75cm]{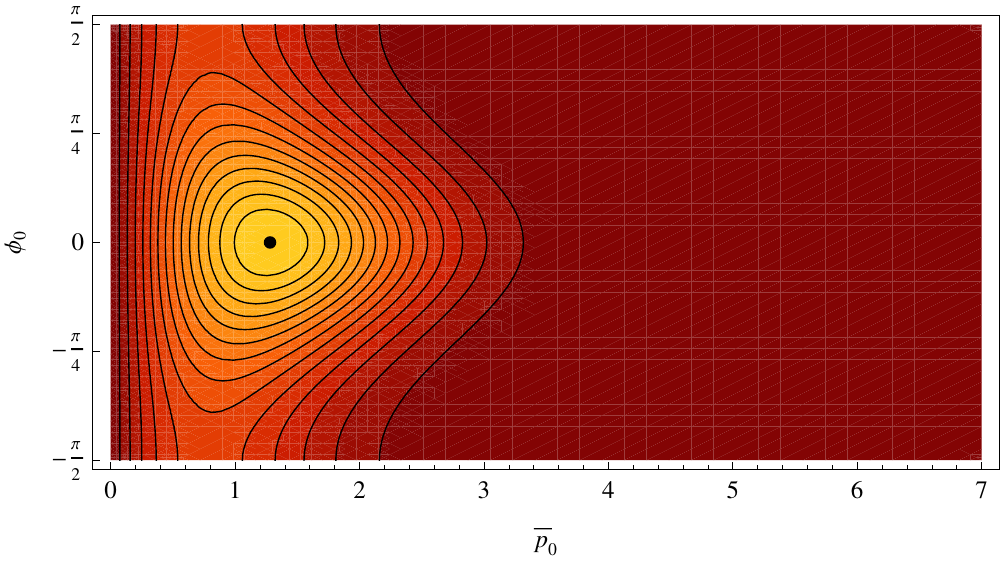}
}
\hspace{1cm}
\subfigure[$p/\sigma{}=2.0$ and $\phi{}=0$.]
{
    \label{fig:Bj_plot_c}
    \includegraphics[width=8.75cm]{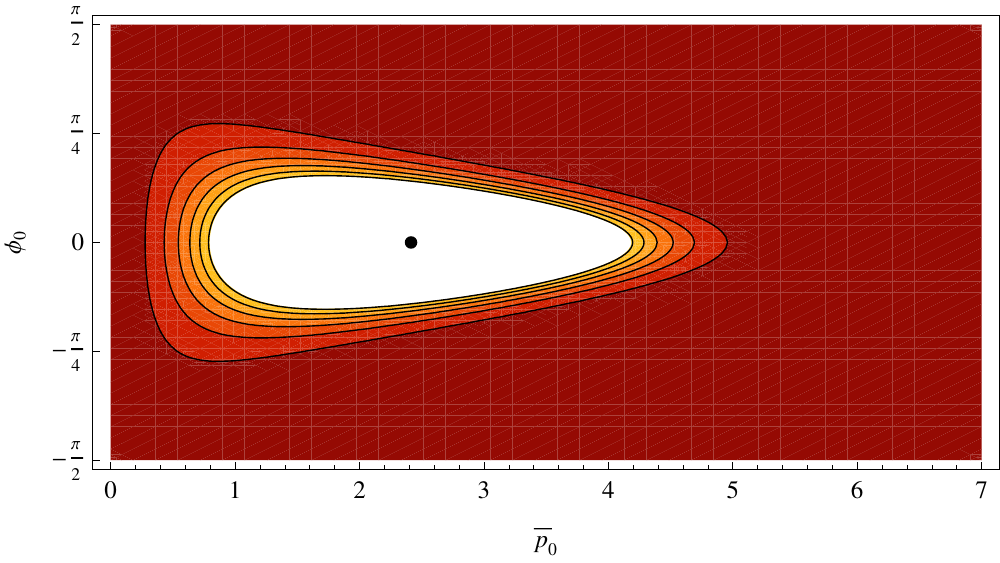}
}
\caption{Contour plots of $\overline{B}(\overline{p},\phi{}|\overline{p}_0,\phi{}_0)$ (Eq.~\ref{eq:posteriorSNR})
for a few different values of $p/\sigma{}$ with $\phi{}=0$  assuming a Jeffreys prior ($\overline{\kappa}(\overline{p}_0,\phi_0)=\sigma^2 2 \overline{p}_0/\pi$) and
$I_0/\Sigma{}=100$.
The dots indicate the maximums of the distributions. Contour spacings are at intervals of $0.02$.
This figure is nearly the same as Fig.~\ref{fig:angleplot} with the roles of $p$ and $p_0$ reversed.
While for small values of $p/\sigma$ and $p_0/\sigma$ the plots appear identical, they are
not. Their similarity diverges for larger values of signal-to-noise, as they must
since $p_0/\sigma$ has a maximum value while $p/\sigma$ does not.}
\label{fig:posteriorJ}
\end{figure}

\begin{figure}[h]
\centering
\subfigure[$p/\sigma{}=0.0$ and $\phi{}=0$.]
{
    \label{fig:posteriora}
    \includegraphics[width=8.75cm]{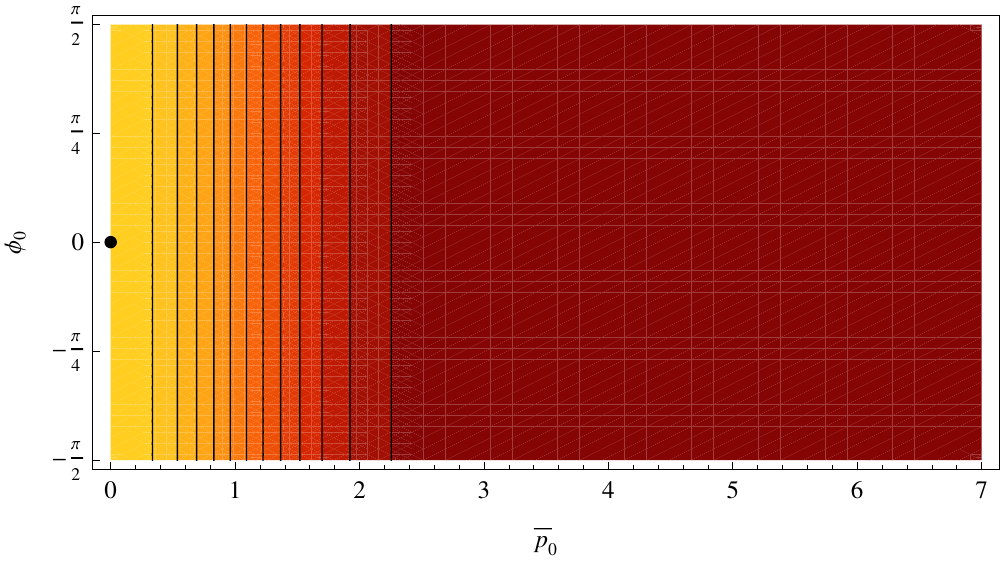}
}
\hspace{1cm}
\subfigure[$p/\sigma{}=0.5$ and $\phi{}=0$.]
{
    \label{fig:posteriorb}
    \includegraphics[width=8.75cm]{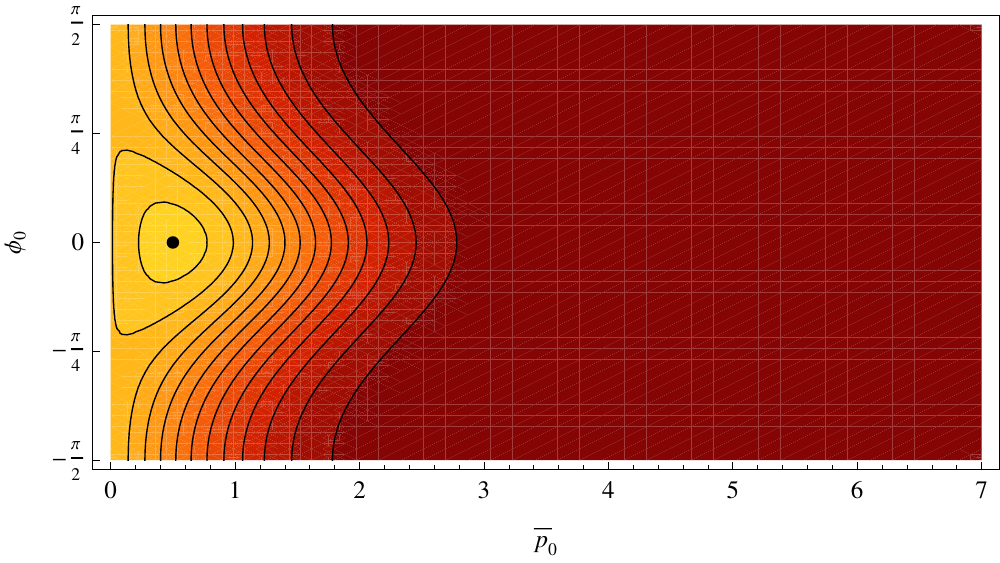}
}
\hspace{1cm}
\subfigure[$p/\sigma{}=2.0$ and $\phi{}=0$.]
{
    \label{fig:posteriorc}
    \includegraphics[width=8.75cm]{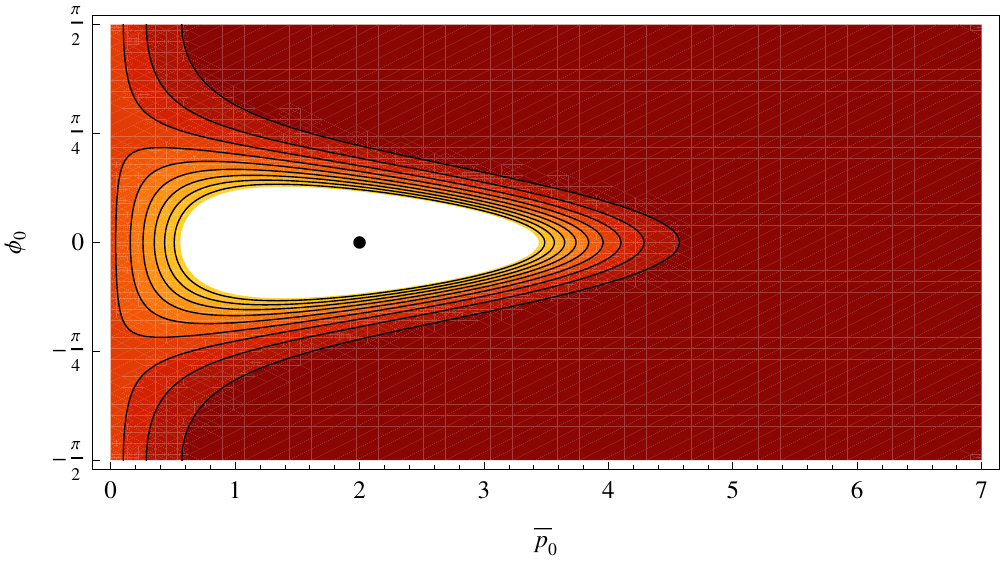}
}
\caption{Contour plots of $\overline{B}(\overline{p}_0,\phi{}_0|\overline{p},\phi{},\sigma)$ (Eq.~\ref{eq:posteriorSNR})
for a few different values of $p/\sigma{}$ with $\phi{}=0$ assuming a uniform polar prior ($\overline{\kappa}(\overline{p}_0,\phi_0)=\sigma/\pi$) and
$I_0/\Sigma{}=100$. The dots indicate the maximums of the distributions. Contour spacings are at intervals of $0.02$.}
\label{fig:posterior}
\end{figure}

Eq.~\ref{eq:posteriorSNR} is shown in Figs.~\ref{fig:posteriorJ} and~\ref{fig:posterior}
for $I_0/\Sigma{}=100$ and two different choices of prior.
Fig.~\ref{fig:posteriorJ} uses $\overline{\kappa}(\overline{p}_0,\phi_0)=\sigma^2 2 \overline{p}_0/\pi$
and Fig.~\ref{fig:posterior} uses $\overline{\kappa}(\overline{p}_0,\phi_0)=\sigma/\pi$.
These priors are discussed in more detail in Section~\ref{Elicitation}. Each plot shows
several different values of $p/\sigma{}$. As before, changing the value
of $\phi$ merely shifts the distribution along the $\phi_0$ axis so $\phi{}=0$ has been used.
In Figs.~\ref{fig:Bj_plot_a} and~\ref{fig:posteriora}, $p/\sigma{}=0$ and
there is no preference for any $\phi{}_0$ over another. In Figs.~\ref{fig:Bj_plot_b}
and~\ref{fig:posteriorb}, $p/\sigma{}=0.5$. Here
a probability bubble is beginning to form. In Figs.~\ref{fig:Bj_plot_c} and~\ref{fig:posteriorc},
$p/\sigma{}=2.0$ and the probability bubble is now fairly mature. The panels
of Fig.~\ref{fig:posterior} are similar. Notice that Fig.~\ref{fig:posteriorJ} is nearly
identical to the distribution plotted in Fig.~\ref{fig:angleplot}. There
is a very weak dependence of $\overline{B}$ on $\sigma$ that 
becomes larger for larger values of $p_0$. If $p_0$ were plotted near the maximum,
the plots would differ radically. When $1/\sigma \rightarrow \infty$,
the plots do become identical. The similarity (and differences)
will be discussed later in conjunction with the one-dimensional, marginalized version
of this plot. 

\subsection{Simple Bayesian Estimators}
Bayesian estimators based on the mean, median, and mode
can be defined for the posterior distribution.

The mode estimator (usually called the MAP or maximum
\textit{a posteriori} estimate) is the maximum of the posterior
distribution. If any delta function is used in the prior, this
causes delta functions to appear in the posterior distribution,
which trumps any finite maximum of the posterior distribution and
renders the mode estimate somewhat useless without further modification.
If no delta functions and a uniform prior is used, then the mode estimate
is the same as the ML estimate (Eq.~\ref{eq:ML}).

The median estimator, $\hat{p}_{0,MED}$, is the value such that
\begin{equation}
\int_{-\pi/2}^{\pi/2} \! \int_{0}^{\hat{p}_{0,MED}} \! B \, dp_0 d\phi{}_0 = \frac{1}{2}.
\label{eq:medestimator}
\end{equation}
This estimator will usually have to be found numerically. A
median estimator $\hat{\phi}_{0,MED}$ is usually easy to
find from symmetry considerations but for our parameter
space must be defined as
\begin{equation}
\int_{0}^{1} \! \int_{\hat{\phi}_{0,MED}}^{\hat{\phi}_{0,MED}+\frac{\pi}{2}} \! B \, d\phi{}_0 dp_0  = \frac{1}{2}    \quad \text{and} \quad      \int_{0}^{1} \! \int_{\hat{\phi}_{0,MED}-\frac{\pi}{2}}^{\hat{\phi}_{0,MED}} \! B \, d\phi{}_0 dp_0 = \frac{1}{2}  .
\label{eq:medestimator2}
\end{equation}
Under special circumstances, $\hat{\phi}_{0,MED}$ may not be unique but after a non-zero measurement
of $p$, it will generally be the case that it is.

The mean estimators are the values
\begin{equation}
\hat{p}_{0,MEAN}= \int_{-\pi/2}^{\pi/2} \! \int_{0}^{1} \! p_0 B \, dp_0 d\phi{}_0
\label{eq:meanestimatorp}
\end{equation}
and
\begin{equation}
\hat{\phi}_{0,MEAN}= \int_{-\pi/2}^{\pi/2} \! \int_{0}^{1} \! \phi_0 B \, dp_0 d\phi{}_0.
\label{eq:meanestimatorangle}
\end{equation}

These three estimators will be different in general.


\subsection{Bayesian Decision Theory}
The posterior distribution contains \textit{all} information about
the relative likelihood of the model parameters. It represents
what has been learned from the observation. For many problems
it is usually desired to summarize the posterior distribution by a few
statistics such as an estimate of the ``best'' value and some
confidence interval (often called ``credible sets'' in this context).
In Bayesian Decision and Estimation Theory determining the ``best'' estimate of the
model parameters requires defining a loss function and a decision
rule (there are many good standard texts offering much more detail such
as \cite{1985bergerbook} and \cite{1994rogerbook}).
The loss function, $L$, assigns a weight to deviations from the
true value such that measured values that are far away from it incur a greater penalty
than those that are closer. The decision rule assigns an
estimate of the true value based upon the data, which will usually be the estimate
that minimizes the expected posterior probable loss.

Deciding upon a loss function is one of the toughest parts of a Bayesian analysis.
The three most common are squared-error loss, absolute deviation loss, and
the ``0-1'' loss. These loss functions have as their solutions for
each variable the mean, median, and mode values of the posterior,
respectively \citep{1994rogerbook}.

The squared-deviation loss function for the polarization problem may be defined as
the squared distance between the true value $(q_0,u_0)$ and the best estimate
$(q_a,u_a)$ under a Euclidean metric (scaled by $\sigma^2$),
\begin{equation}
L(q_0,u_0,q_a,u_a)= ((q_0-q_a)^2 + (u_0-u_a)^2)/\sigma{}^2.
\label{eq:L}
\end{equation}
Transforming Eq.~\ref{eq:L} to polar coordinates gives,
\begin{equation}
L(p_0,\phi{}_0,p_a,\phi{}_a) = (p_0^2 + p_a^2 - 2 p_0 p_a \cos{}(2(\phi{}_0-\phi{}_a)))/\sigma{}^2.
\end{equation}
The posterior expected loss, $Z$, under $B$ is
\begin{equation}
Z(p_a,\phi_a) = \int_{-\frac{\pi}{2}}^{\frac{\pi}{2}} \! \int_{0}^{1} \! L(p_0,\phi{}_0,p_a,\phi{}_a) B(p_0,\phi{}_0|p,\phi{},\sigma) \, dp_0 d\phi_0.
\end{equation}
Loss can be minimized by solving $\frac{\partial{}Z}{\partial p_a}=0$ and $\frac{\partial{}Z}{\partial \phi{}_a}=0$
and checking that the solution is a minimum via the second derivative test. These two 
equations are difficult to solve completely but it can be shown that if
$\kappa{}(p_0,\phi_0) = \kappa{}(p_0)$ then
\begin{equation}
\hat{\phi}_a=\phi_0=\phi
\end{equation}
and
\begin{equation}
\hat{p}_a = \int_{-\frac{\pi}{2}}^{\frac{\pi}{2}} \! \int_{0}^{1} \! p_0 B(p_0,\phi{}_0|p,\phi{},\sigma)   \, dp_0 d\phi{}_0
\end{equation}
are a solution and
$\frac{\partial^2 Z}{\partial p_a^2}  + 
\frac{\partial^2 Z}{\partial \phi{}_a^2} -
\left(\frac{\partial^2 Z}{\partial p_a \partial \phi{}_a}\right)^2 > 0$
when $p_a,p_0>0$ so the solutions are a minimum.

The absolute loss function is
\begin{equation}
L(q_0,u_0,q_a,u_a)= \sqrt{(q_0-q_a)^2 + (u_0-u_a)^2}/\sigma{}.
\end{equation}
It will be assumed that the usual median estimators in polar
coordinates are in fact a solution to this loss function.

The ``0-1'' loss function is
\begin{equation}
L(q_0,u_0,q_a,u_a) =  \begin{cases}
1 & \text{if} \quad q_a=q_0 \,\, \text{and} \,\, u_a=u_0 \\
0 & \text{if} \quad q_a \ne q_0 \,\, \text{and} \,\, u_a \ne u_0
\end{cases}.
\end{equation}
It will also be assumed that the mode is the solution of this
loss function. When the prior contains a delta function,
this estimator is not very useful.

\subsection{Posterior odds}
An analog of hypothesis testing in Bayesian analysis is
the comparison of posterior odds. In the simplest case,
suppose one wishes to test the hypothesis that
the model parameters lie in some subset $\Theta_1$ of the
parameter space. This is accomplished by integrating the
posterior distribution over the subset. If the probability
is greater than $50\%$, the hypothesis is accepted. If it
is less than $50\%$, it is rejected. The simplicity of
these tests is one of the benefits of the Bayesian approach.
The subset need not consist of more than a single point
but in that case the associated probability will usually be infinitesimal
unless distributions are allowed for priors.
In the next section, priors will be introduced that use
a delta function at the origin to represent the probability
that a source has negligible polarization. After the measurement,
one may compute the posterior odds for the origin (a single
point) to test if the measurement is consistent with zero.
One of the benefits of $\hat{p}_{0,MED}$ is that it naturally
performs this test for a delta function at the origin.

Two probabilities are of interest, the probability that $p_0$ is zero and the
probability that $p_0$ is greater than zero. In mathematical form,
\begin{equation}
\text{Prob}(p_0=0) = \lim_{\alpha \to 0^{+}} \int_{-\pi/2}^{\pi/2} \! \int_{0}^{\alpha{}} \! B(p_0,\phi_0|p,\phi{},\sigma{}) \, dp_0 d\phi{}_0
\label{probzero}
\end{equation}
and
\begin{equation}
\text{Prob}(p_0 > 0) = \lim_{\alpha \to 0^{+}} \int_{-\pi/2}^{\pi/2} \! \int_{\alpha{}}^{1} \! B(p_0,\phi_0|p,\phi{},\sigma{}) \, dp_0 d\phi{}_0.
\end{equation}
Since the distribution is normalized, only Eq.~\ref{probzero} need actually be calculated
because $\text{Prob}(p_0=0) + \text{Prob}(p_0 > 0) = 1$.
If $\text{Prob}(p_0=0) > 1/2$, then it is more likely the source has negligible than
appreciable polarization. If $\text{Prob}(p_0=0) < 1/2$, then it is more likely the source has appreciable than
negligible polarization.

It is now time to discuss choosing a prior so that these equations may
be used.

\section{Elicitation of priors}
\label{Elicitation}
In Bayesian statistics, the choice of the prior distribution is non-trivial
and can be very controversial because it
is frequently subjective and based on previous experience for
the problem at hand. The purpose of the prior density is to
represent knowledge of the model parameters \textit{before} an experiment.
One common situation is where one has no prior knowledge of
their values and wishes to construct a prior to represent this ignorance.
Such priors are called ``objective''. Constructing objective priors can be extremely
subtle. The Bertrand Paradox illustrates that the method of producing or
defining a ``random'' value can have important consequences for the resulting
probability distributions \citep{1889Bertrandbook}. In general, the solution to the Bertrand Paradox
is that ``real randomness'' requires certain translation and scaling invariance
properties \citep{1973FoPh....3..477J,1984Tissier,2010arXiv1008.1878D}.
In practice, such objective priors may contain too little
information to satisfactorily model the phenomena being studied.

In general, the polarization state of an unresolved source at a particular wavelength
is described by a point of the Poincar\'{e} sphere, which is the
three-dimensional set of the physically possible values for the $q$, $u$, and $v$ Stokes parameters
given an intensity. An assumption was made in the introduction that circular polarization
is negligible. This is a \textit{physical} assumption which influences
the meaning of the polarization state of a ``random'' source. The resulting $q$-$u$ distribution
of random sources is not to be confused with the expected $q$-$u$ distribution of
a random source whose $v$ component was ignored. For a uniform distribution of
points within the Poincar\'{e} sphere, this latter distribution would be the density distribution
obtained by projection onto the $q$-$u$ plane, which would have a larger density
near the origin. This distribution is important and interesting but its three-dimensional motivation
is somewhat outside two-dimensional scope of this work. A brief discussion of it
is given in Appendix~\ref{projappend}.

Investigation is still required into the meaning of ``randomness'' even under just
the $V=0$ point-of-view. It is necessary to understand what role the choice of coordinates plays
and how ``uniform'' does not imply ``random''.

\subsection{Uniform prior in Cartesian coordinates (the Jeffreys prior)}
\begin{figure}
\centering
\subfigure[A set of $2000$ points generated using the Jeffreys prior.]
{
    \label{fig:unitdiska}
    \includegraphics[width=6cm]{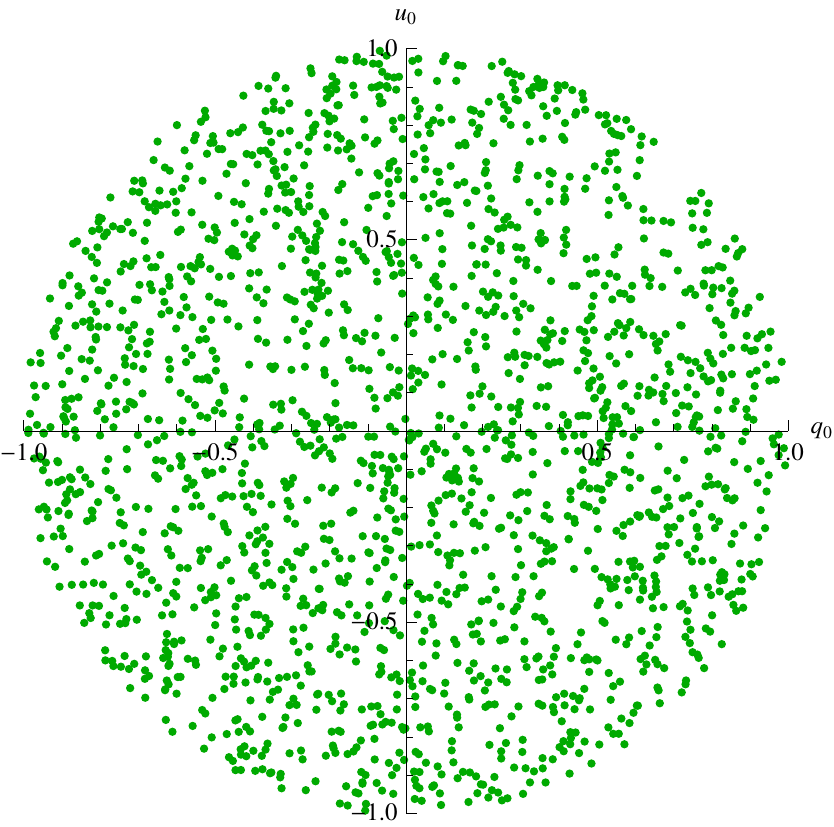}
}
\hspace{1cm}
\subfigure[A second set of $2000$ points in $p_0$-$\phi{}_0$ space generated for the uniform polar prior.]
{
    \label{fig:unitdiskb}
    \includegraphics[width=6cm]{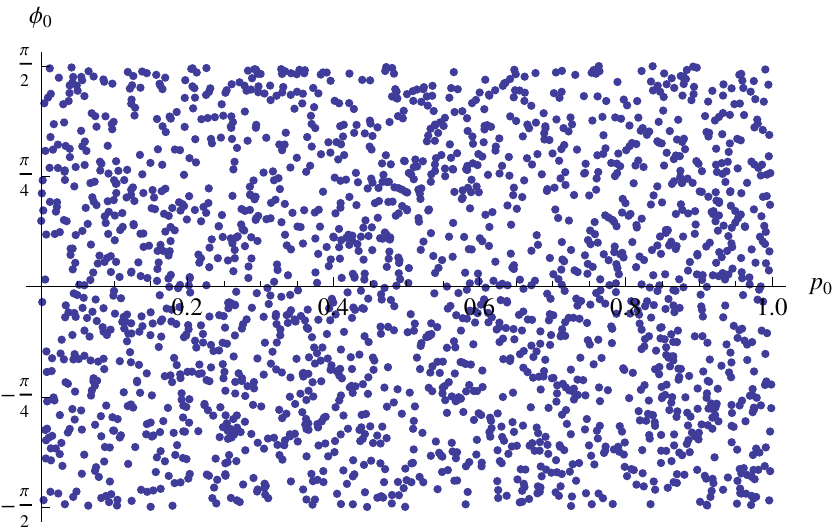}
}
\hspace{1cm}
\subfigure[The same set as in Fig.~\ref{fig:unitdiskb} in $q_0$-$u_0$ space.]
{
    \label{fig:unitdiskc}
    \includegraphics[width=6cm]{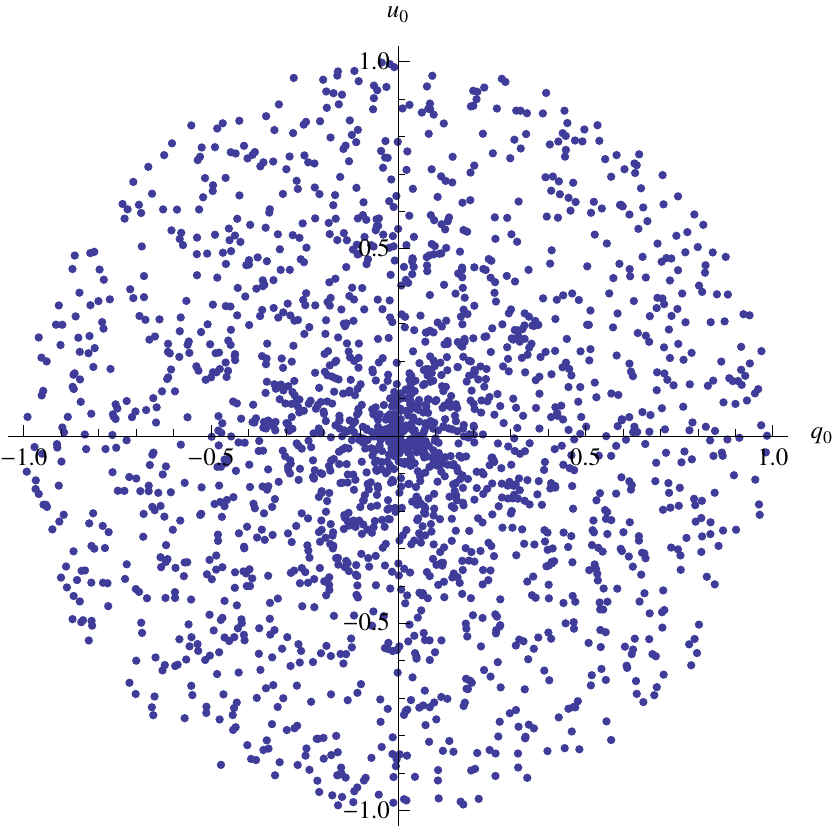}
}
\caption{A comparison of ``random'' points generated with
the Jeffreys prior and the uniform polar prior.}
\label{fig:unitdisk}
\end{figure}

The prior most likely to coincide with a person's intuitive notion
of a ``random point'' in the unit disk is a uniform distribution
in $q_0$-$u_0$ coordinates is
\begin{equation}
\kappa{}_C(q_0,u_0)=\frac{1}{\pi}.
\label{uniformpriorqu}
\end{equation}
In polar coordinates, this is equivalent to
\begin{equation}
\kappa{}(p_0,\phi{}_0)=\frac{2 p_0}{\pi} \quad \quad \text{or} \quad \quad \overline{\kappa}(\overline{p}_0,\phi{}_0)=\sigma^2\frac{2 \overline{p}_0}{\pi}.
\label{polarbarred}
\end{equation}
The barred versions are always $\overline{\kappa}(\overline{p}_0,\phi_0)= \sigma \kappa(\sigma{} \overline{p}_0,\phi_0)$.
Both distributions are normalized. Random points generated
using this distribution have an expected density that is constant across the
unit disk (see Fig.~\ref{fig:unitdiska}). One aspect of this
prior is that a randomly generated point will generally be more likely to
have a strong polarization than a weak one because an annulus
associated with a large radius has more area than for a smaller radius.

It can be shown that the prior given by Eq.~\ref{uniformpriorqu} is
the normalized Jeffreys prior for the polarization problem.
The Jeffreys prior is an objective prior constructed from the
square root of the determinate of the Fisher information matrix \citep{Jeffreys1939theory,1946Jeffries,2008arXiv0804.3173R,2009arXiv0909.1008R}.
One of the features of the Jeffreys prior is that it is invariant under
a re-parametrization. As an objective
prior, the Jeffreys prior is in some sense the prior desired when
absolutely no previous knowledge of the model parameters is known.

While the objectivity of this prior is admirable, in practice,
most astronomical sources are not expected to have very large polarizations
(masers can be a notable exception); so, in many cases, it is undesirable
that the polarization be more likely to be large than small.
In this respect, this prior may not be a good choice for some
astrophysical observations despite its objectivity: it
simply contains too little information to be suitable
and overestimates our lack of knowledge and previous experience.

\subsection{Uniform prior in polar coordinates}
The uniform prior in polar coordinates captures some features
that are usually desired for astronomical sources than the
previous case. Its (normalized) formula is
\begin{equation}
\kappa{}(p_0,\phi{}_0)=\frac{1}{\pi} \quad \quad \text{or} \quad \quad  \overline{\kappa}(\overline{p}_0,\phi{}_0)=\frac{\sigma}{\pi}.
\label{uniformprior}
\end{equation}
While this prior may be ``uniform''
in $p_0$-$\phi_0$ space (see Fig.~\ref{fig:unitdiskb}), it is definitely not uniform in $q_0$-$u_0$
space. In fact, in that space, the density of points produced randomly by this prior increases
near the origin, as is shown in Fig.~\ref{fig:unitdiskc}. This is a subjective prior
in the sense that it prefers points closer to the origin.

\subsection{The addition of a delta term at the origin}
One subtle feature of the previous priors is that they do not
allow for a finite probability that an object's
polarization is exactly zero. Instead there is only a finite
chance that the polarization will be in some interval
that includes zero. A Dirac delta function at the origin may be
used to allow for the possibility that an object
truly has zero polarization. ``Exactly zero'' polarization may be interpreted
as equivalent to the statement that there is a large probability that the polarization is negligible.
In this case, the delta function then becomes just a mathematical tool to replace
a probability density that rises extremely sharply at the origin in such a way to bound
an appreciable probability.

When one is investigating \textit{if} some class of objects are polarized,
it seems reasonable to assign a 50\% chance to the probability that the object
is unpolarized and a 50\% chance to it being polarized. More generally, the
percentage can be parametrized, such that the probability of zero polarization
is $A$ and greater than zero polarization is $(1-A)$.
Let the prior consists of two components: a normalized delta function term at the origin,
$\kappa{}_1(p_0,\phi{}_0)$, and a
normalized non-delta function term, $\kappa{}_2(p_0,\phi{}_0)$.
The general form for the prior can now be written as
\begin{equation}
\kappa{}(p_0,\phi{}_0)=A \kappa{}_1(p_0,\phi{}_0) + (1-A) \kappa{}_2(p_0,\phi{}_0)
\label{generalprior}
\end{equation}
or
\begin{equation}
\overline{\kappa}(\overline{p}_0,\phi{}_0)=A \overline{\kappa}_1(\overline{p}_0,\phi{}_0) + (1-A) \overline{\kappa}_2(\overline{p}_0,\phi{}_0).
\label{generalprior}
\end{equation}
The delta term must be
\begin{equation}
\kappa{}_1(p_0,\phi{}_0)=\frac{2 \delta{}(p_0)}{\pi} \quad \text{or} \quad \overline{\kappa}_1(\overline{p}_0,\phi{}_0)=\frac{2 \delta{}(\overline{p}_0)}{\pi},
\label{deltaterm}
\end{equation}
respectively. The two versions in Eq.~\ref{deltaterm} have the same form because the delta function transforms in a way that
cancels the $\sigma$'s that typically will appear in the barred version.
Notice too that the $p_0$-delta function occurs at the lower limit of the $p_0$ integration.
This is a somewhat quirky situation but by remembering that delta functions are
constructed as a limit of a sequences of functions it is easily seen that
$\int_0^{\alpha{}} \! f(x)\delta{}(x) \, dx=f(0)/2$ for $\alpha{}>0$. Taking into account the factors
of one-half that arise in this way, it can be shown that Eqs.~\ref{generalprior} and~\ref{deltaterm} are
both normalized properly.\footnote{This
idea of delta functions on boundaries (or even boundaries of boundaries) helps motivate and
generally requires the concept of Lebesgue density. Extensions of this paper to include circular
polarization, $V$, would have to be careful to get the Lebesgue density factors correct.} A
delta function could also be placed at any other point in the disk of physically possible values but the
origin is unique as it is the only case that preserves rotational symmetry.

The previous priors with the addition of a delta term become
\begin{equation}
\overline{\kappa}(\overline{p}_0,\phi{}_0)=\frac{2 A \delta{}(\overline{p}_0)}{\pi} + (1-A)\sigma{}^2\frac{2 \overline{p}_0}{\pi} \quad \text{(Jeffreys)}
\end{equation}
and
\begin{equation}
\overline{\kappa}(\overline{p}_0,\phi{}_0)=\frac{2 A \delta{}(\overline{p}_0)}{\pi} + \sigma{}\frac{1-A}{\pi} \quad \text{(uniform polar)}.
\label{eq:barredprior}
\end{equation}
These priors simplify to the more typical cases when $A=0$. The delta
term however is an interesting addition. It is only with the addition of the delta term
and the use of the median estimator that one is allowed to state the source is
consistent with no polarization from the posterior distribution.
It seems most reasonable to assign $A$ the value $0.5$ when it
is unknown if the source is polarized. Another interesting
choice is $A=\frac{1}{1+\pi}$. This value causes the $\text{Prob}(p_0=0)$ for
the prior to be equal to the probability density of the non-delta function component.
There is a temptation to say that it recaptures some translation
invariance properties that are otherwise lost upon the introduction
of the delta term. It is unclear to the author if this has any
mathematical significance or indeed how to justify a particular
choice of $A$ in general. Using $A=0$ is the most conservative choice
and is probably the most appropriate in many experiments unless a
non-zero $A$ is explicitly required.
A non-zero value of $A$, as will be seen, can cause the median estimator of $\overline{p}_0$
to be zero below some critical value of $\overline{p}$. One way of interpreting this
is that the median estimator determines what value of polarization must be measured to
convince a person who is skeptical of polarization with degree $A$ that a source is
actually polarized. Ultimately, it is up to the researcher to decide the best prior for the
experiment.

Under Eq.~\ref{generalprior}, the posterior distribution reduces to
\begin{equation}
\overline{B}(\overline{p}_0,\phi_0|\overline{p},\phi{},\sigma) =
\frac{A \, \overline{f} \, \overline{\kappa}_1 + (1-A) \, \overline{f} \, \overline{\kappa}_2}
{A \frac{\overline{p}}{\pi} \exp \left( -\frac{\overline{p}^2}{2}\right) + (1-A) \int_{-\pi/2}^{\pi/2} \! \int_0^{1/\sigma} \! \overline{f} \, \overline{\kappa}_2    \, d\overline{p}'_0 d\phi'_0  }.
\end{equation}
Figs.~\ref{fig:posteriorJ} and~\ref{fig:posterior} were special cases of this formula.

\subsection{Other constructions}
Bayesian statisticians have invented a number of other techniques
for constructing priors such as using conjugate families and building
hierarchical models. These are not covered.

\section{Marginal distributions of polarization magnitude}
The marginal posterior distribution, $S$, of $p_0$ over $B$ is given by
\begin{equation}
S(p_0|p,\phi{},\sigma) = \int_{-\pi{}/2}^{\pi{}/2} \! B(p_0,\phi{}_0|p,\phi{},\sigma)    \,   d\phi{}_0.
\label{eq:S}
\end{equation}
The denominator of $B$ is a function of $p$ and $\phi{}$ only and 
may be pulled through the integral. The numerator becomes
$\int_{-\pi/2}^{\pi/2} \! f(p,\phi{}|p_0,\phi{}_0,\sigma{}) \kappa{}(p_0,\phi{}_0) \, d\phi_0$.
No further simplifications can be made unless a prior is chosen.
While not necessary, it will frequently be the
case that priors for a first measurement are independent of angle, that is,
$\kappa{}(p_0,\phi_0)=\kappa{}(p_0)$. If so, then it is easy to show
that
\begin{equation}
S(p_0|p,\sigma) = \frac{ R(p|p_0,\sigma{})  \kappa{}(p_0)    }
{  \int_{0}^{1} \! R(p|p'_0,\sigma{})   \kappa{}(p'_0)      \,  dp'_0     }.
\label{returnS}
\end{equation}
(An integration of the sampling function under consideration, $f(p,\phi{}|p_0,\phi{}_0,\sigma{})$,
arrives at the same result, Eq.~\ref{eq:rice}, regardless if the integration
is over $\phi_0$ or $\phi$ due to the symmetry in those variables.)
The resulting new posterior, which may be treated as a prior for the next measurement,
will not, in general, be independent of angle. The condition $\kappa{}(p_0,\phi{}_0)=\kappa{}(p_0)$
therefore does not extend to multiple measurements. Those are not the focus
of this investigation and none of the $\kappa$-priors examined so far vary with
angle. Eq.~\ref{returnS} turns out to be exactly the same form as the posterior Rice distribution,
which is now examined.

\subsection{Posterior Rice distributions}
\citet{2006PASP..118.1340V} uses a Bayesian approach to argue for the use of the posterior
Rice distribution to calculate the maximum likelihood and error bars on measured
polarization. This special case was the motivation for the more general theory of this
paper.

Introduce new one-dimensional prior $\tau{}(p_0)$.
The posterior Rice distribution, $\rho{}$, is
\begin{equation}
\rho{}(p_0|p,\sigma{}) = \frac{R(p|p_0,\sigma{}) \tau{}(p_0) }{  \int_{0}^{1} \! R(p|p'_0,\sigma{}) \tau{}(p'_0) \, dp'_0     }.
\label{eq:vai_rho}
\end{equation}
The conversion to barred priors is given
by $\overline{\tau}(\overline{p}_0)=\sigma \tau{}(\sigma{} \overline{p}_0)$.
The barred version of the formula is
\begin{equation}
\overline{\rho{}}(\overline{p}_0|\overline{p},\sigma) = \frac{\overline{R}(\overline{p}|\overline{p}_0) \overline{\tau}{}(\overline{p}_0) }{      \int_{0}^{1/\sigma{}} \! \overline{R}(\overline{p}|\overline{p}'_0) \overline{\tau}{}(\overline{p}'_0) \, d\overline{p}'_0     }.
\label{eq:vai_rhoSNR}
\end{equation}

The objective prior in one dimension is $\tau(p_0)=1$ so
technically the one-dimensional Jeffreys prior and the uniform polar prior are the same.
As can be seen from Eq.~\ref{returnS} and Eq.~\ref{eq:vai_rho}, when
$\kappa(p_0)$ is independent of angle, $S$ and $\rho$ have the same form with
$\kappa(p_0)$ playing the role of $\tau(p_0)$. The ``one-dimensional'' version of the two-dimensional Jeffreys prior
is $\tau(p_0)=2 p_0$ and the one-dimensional version of the two-dimensional
uniform polar prior is $\tau(p_0)=1$. There is only an immaterial
factor of $1/\pi$ difference between $\kappa(p_0)$
and $\tau(p_0)$. In a slight abuse of terminology, $\tau(p_0)=2 p_0$
will still be called the ``Jeffreys prior'' despite switching to the one-dimensional notation.
The two-dimensional theory is still within context because of Eq.~\ref{returnS}.
In barred variables, the priors are $\overline{\tau}(\overline{p}_0)=\sigma{}^2 2 \overline{p}_0$
and  $\overline{\tau}(\overline{p}_0)=\sigma{}$, respectively.

\subsubsection{Mean estimator for the posterior Rice distribution}
The mean estimator, $\hat{\overline{p}}_{0,MEAN}$, is given by
\begin{equation}
\hat{\overline{p}}_{0,MEAN} = \int_0^{1/\sigma} \! \overline{p}_0 \, \overline{\rho}(\overline{p}_0|\overline{p},\sigma) \, d\overline{p}_0.
\label{onedmeansimp}
\end{equation}
Its value is always greater than zero for any given $\overline{p}$ and will
usually have to be found numerically.

\subsubsection{Median estimator for the posterior Rice distribution}
The median estimator, $\hat{\overline{p}}_{0,MED}$, for the posterior Rice distribution
is the value of the upper integrand such that
\begin{equation}
\int_0^{\hat{\overline{p}}_{0,MED}} \! \overline{\rho{}}(\overline{p}_0|\overline{p},\sigma) \, d\overline{p}_0 = \frac{1}{2}.
\label{onedmedian}
\end{equation}
This expression will also have to be solved numerically in most cases.

\subsection{Selected plots of the posterior Rice distribution}
\begin{figure}
\subfigure[This plot uses $I_0/\Sigma{}=100$. The blue dotted line
is the median estimator. The black dashed line is the mean estimator.]
{
    \label{fig:jeff_contour_rho}
    \includegraphics[width=8.5cm]{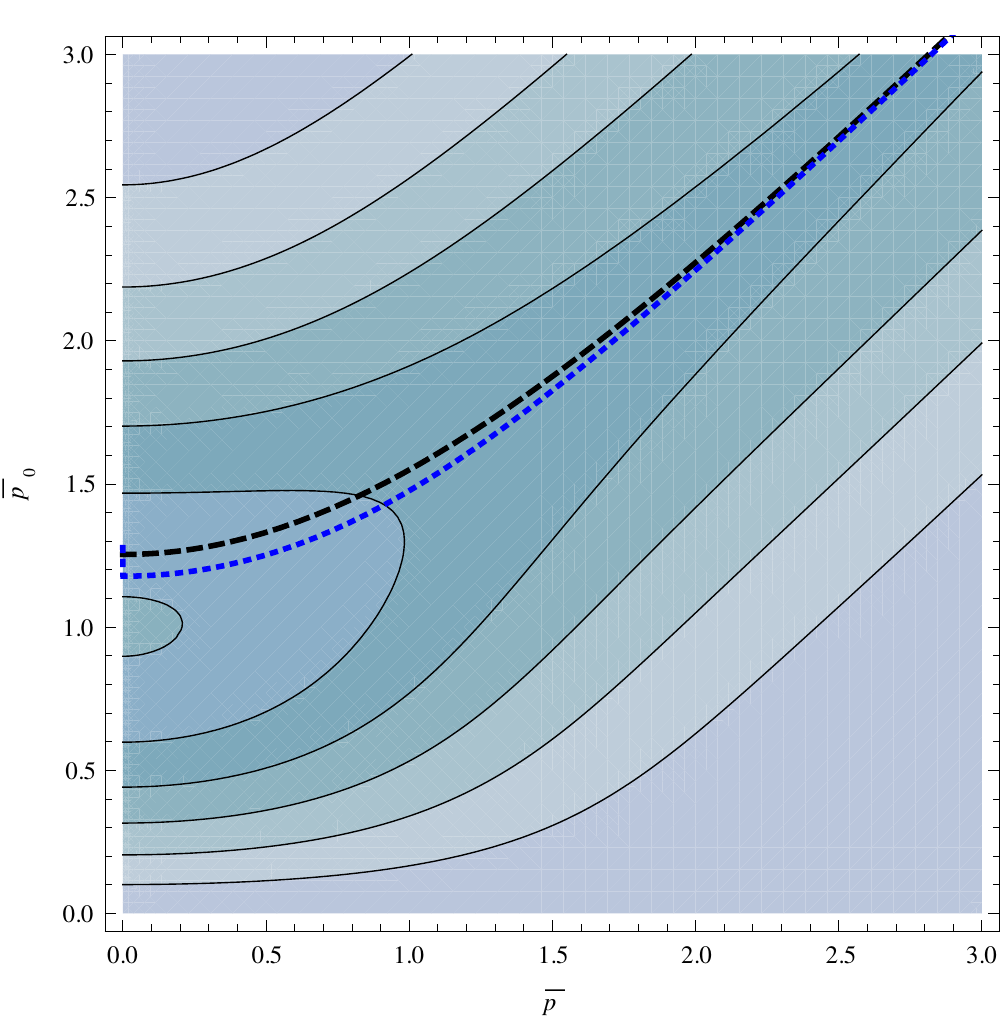}
}
\hspace{0.25cm}
\subfigure[This plot uses $I_0/\Sigma{}=8$ (a very small value) and shows the behavior
for $\overline{p} > I_0/\Sigma{}$. It is clearly seen that for large $\overline{p}$,
there is a large probability that the value of $\overline{p}_0$ is near the maximum ($8$ in this case).
Fig.~\ref{fig:jeff_contour_rho} would exhibit similar behavior in its large polarization
regime near and past $\overline{p}=100$.]
{
    \label{fig:jeff_contour_rho_finite}
    \includegraphics[width=8.5cm]{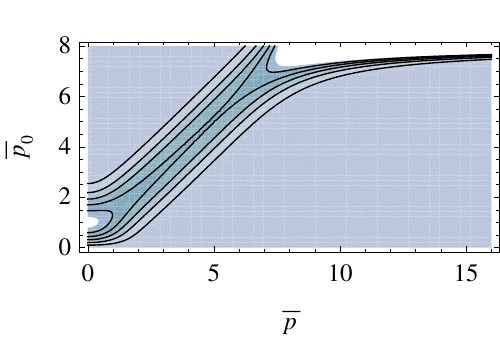}
}
\caption{Two contour plots of the posterior Rice distribution (Eq.~\ref{eq:vai_rhoSNR})
with $\overline{\tau}(\overline{p}_0)=\sigma^2 2 \overline{p}_0$ and two values of $\sigma$. The contours are spaced by $0.1$ intervals. 
}
\label{fig:jeff_contour_two_panel}
\end{figure}

Fig.~\ref{fig:jeff_contour_two_panel} shows two versions of the posterior
Rice distribution using the Jeffreys prior, $\overline{\tau}(\overline{p}_0)=\sigma{}^2 2 \overline{p}_0$.
Fig.~\ref{fig:jeff_contour_rho} shows
the distribution near the origin and assuming $1/\sigma=100$. Fig.~\ref{fig:jeff_contour_rho_finite}
shows the full range of $\overline{p}_0$ assuming $1/\sigma=8$. This figure
exhibits several key results. For the Jeffreys prior, the posterior Rice
distribution is similar to just the transpose of the Rice distribution.
This similarity breaks down at values of $\overline{p}$ near or exceeding
the maximum value of $\overline{p}_0$. The mean (black) and median (blue)
estimator curves, are very similar. They are also similar to the (transposed) one-dimensional
Rice distribution estimator curves but that similarity also breaks down at near the maximum
value of $\overline{p}_0$.

\begin{figure}
\subfigure[This plot uses $I_0/\Sigma{}=100$.  The blue dotted line
is the median estimator. The black dashed line is the mean estimator.]
{
    \label{fig:contour_rho}
    \includegraphics[width=8.5cm]{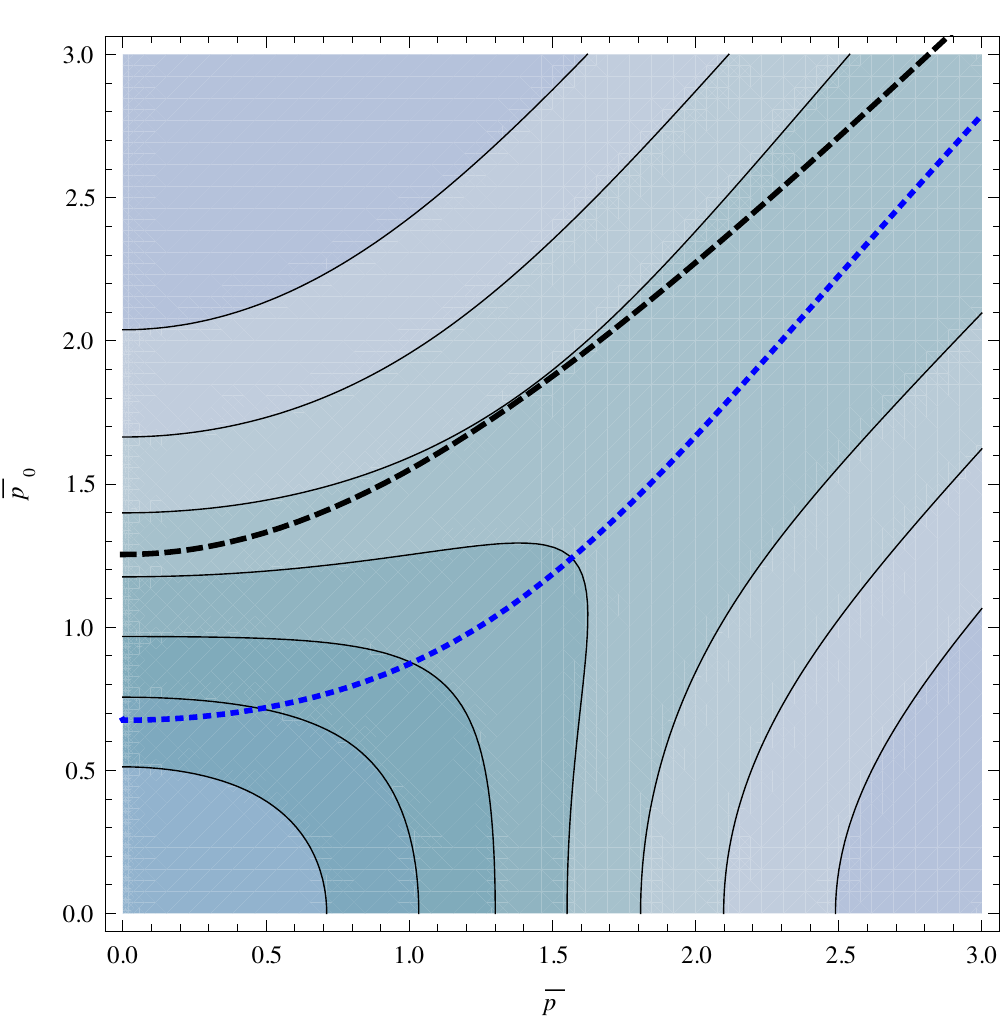}
}
\hspace{0.25cm}
\subfigure[This plot uses $I_0/\Sigma{}=8$ (a very small value) showing the behavior
for $\overline{p} > I_0/\Sigma{}$. It is clearly seen that for large $\overline{p}$,
there is a large probability that the value of $\overline{p}_0$ is near the maximum
($8$ in this case). Fig.~\ref{fig:contour_rho} would exhibit similar behavior
in its large polarization regime near and past $\overline{p}=100$.]
{
    \label{fig:contour_rho_finite}
    \includegraphics[width=8.5cm]{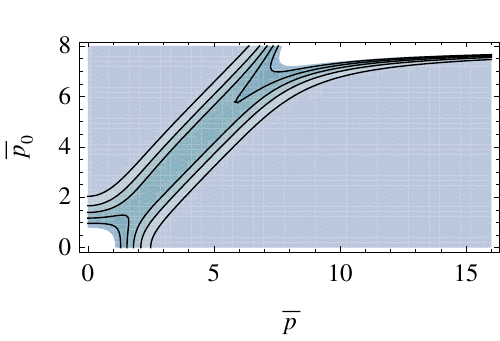}
}
\caption{Two contour plots of the posterior Rice distribution (Eq.~\ref{eq:vai_rhoSNR}) with
$\overline{\tau}(\overline{p}_0)=\sigma$ and two values of $\sigma$. The contours are spaced by $0.1$ intervals.
}
\label{fig:contour_rho_two_panel}
\end{figure}

Fig.~\ref{fig:contour_rho_two_panel} uses $\overline{\tau}(\overline{p}_0)=\sigma{}$ and
two different values of $\sigma$.
Fig.~\ref{fig:contour_rho} is nearly the same plot as presented in \citet{2006PASP..118.1340V},
which used $\tau(p_0)=1$ and the unstated assumption
that $I_0/\Sigma{} \to{} \infty$, which allows direct integration of the denominator of Eq.~\ref{eq:vai_rhoSNR}
at the expense of having an unnormalizable prior.
While this approximation will be acceptable most of the time when the signal-to-noise
is large, it is unphysical. 
The true value of $\overline{p}_0$ cannot exceed $1/\sigma{}$ (=$I_0/\Sigma{}$), which is
finite. This is best seen by plotting the full range of $\overline{p}_0$ as
is done in Fig.~\ref{fig:contour_rho_finite} for $I_0/\Sigma{}=8$ (a small value). There it is seen that for
values of $\overline{p}$ that are near or greater than the maximum possible
value of $\overline{p}_0$ ($8$ in this case), the probability density function ``crowds up'' near the maximum value.
This time the mean (black) and median (blue) curves are rather different. The median curve prefers value closer to
$\overline{p}_0=0$. The median estimator seems to be a better statistic in this case as
the mean estimator curve is affected too much by outlier possibility.

A delta term may also be added to the prior as was previously motivated.
Let $\tau{}(p_0)= A \tau_1(p_0) + (1-A) \tau_2(p_0)$, where $\tau_1(p_0)$
is normalized a delta term at the origin and $\tau_2(p_0)$ is a normalized
component without any delta terms. The normalized barred prior is
$\overline{\tau}{}(\overline{p}_0) = A \overline{\tau}_1(\overline{p}_0) + (1-A)\overline{\tau}_2(\overline{p}_0)$.
This time $\tau_1(p_0)=2 \delta{}(p_0)$ and $\overline{\tau}_1(\overline{p}_0)=2 \delta{}(\overline{p}_0)$.
Under this prior, Eq.~\ref{eq:vai_rhoSNR} reduces to
\begin{equation}
\overline{\rho{}}(\overline{p}_0|\overline{p},\sigma) =
\frac{    A \, \overline{R} \, \overline{\tau}_1{}  +  (1-A) \, \overline{R} \, \overline{\tau}_2 }
     {    A \, \overline{p} \, \operatorname{exp} \left(-\frac{\overline{p}^2}{2}  \right)  +  (1-A) \int_0^{1/\sigma} \! \overline{R} \, \overline{\tau}_2  \, d\overline{p}'_0 }.
\label{rho_very_general}
\end{equation}
Here, the function arguments have been dropped for space.
For the two-component prior, the definition of the median estimator (Eq.~\ref{onedmedian})
leads to the following expression
\begin{equation}
\int_0^{\hat{\overline{p}}_{0,MED}} \! \overline{R} \, \overline{\tau}_2 \, d\overline{p}_0 =
      \frac{1}{2} \int_0^{1/\sigma} \! \overline{R} \, \overline{\tau}_2  \, d\overline{p}_0
      - \frac{A}{2(1-A)} \overline{p} e^{-\overline{p}^2/2}.
\label{onedmediansimp}
\end{equation}

\begin{figure}
\centering
\subfigure
{
    \label{fourpaneljeffa}
    \includegraphics[width=4.1cm]{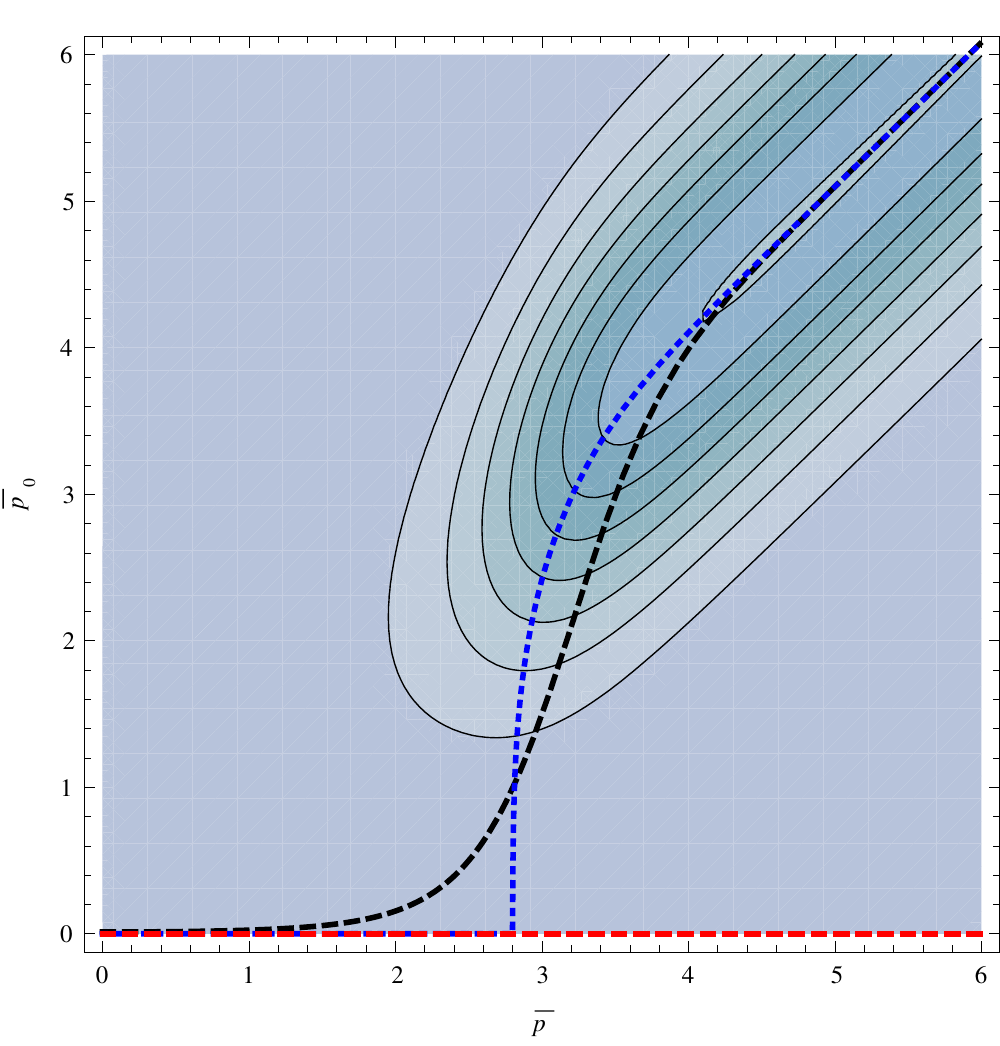}
}
\hspace{0.25cm}
\subfigure
{
    \label{fourpaneljeffb}
    \includegraphics[width=4.1cm]{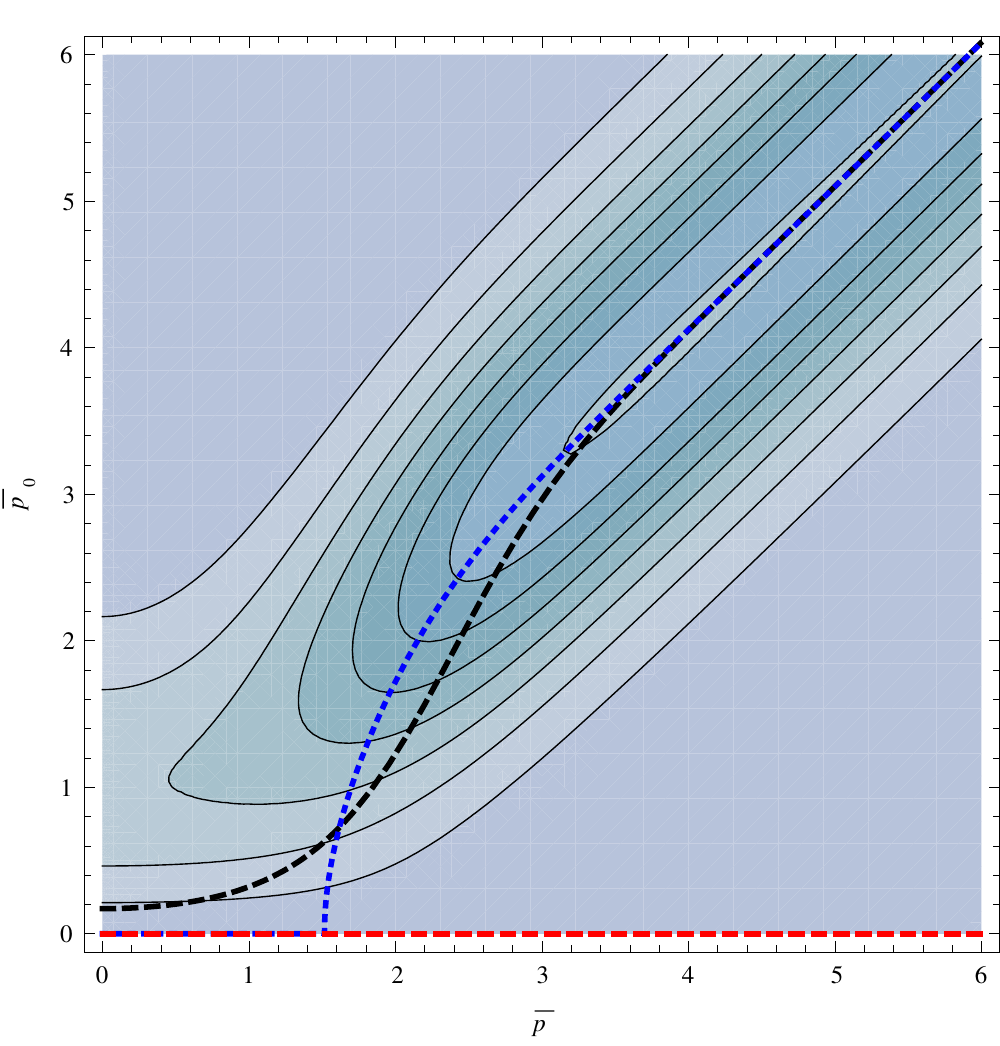}
}
\addtocounter{subfigure}{-1}
\addtocounter{subfigure}{-1}
\hspace{0.25cm}
\subfigure[$A=0.01$ and $\sigma{}=0.01$]
{
    \label{fourpaneljeffaunder}
    \includegraphics[width=4.1cm]{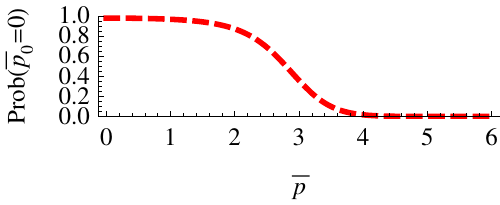}
}
\hspace{0.25cm}
\subfigure[$A=0.01$ and $\sigma{}=0.04$]
{
    \label{fourpaneljeffbunder}
    \includegraphics[width=4.1cm]{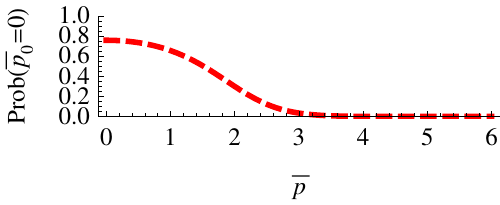}
}
\hspace{0.25cm}
\subfigure
{
    \label{fourpaneljeffc}
    \includegraphics[width=4.1cm]{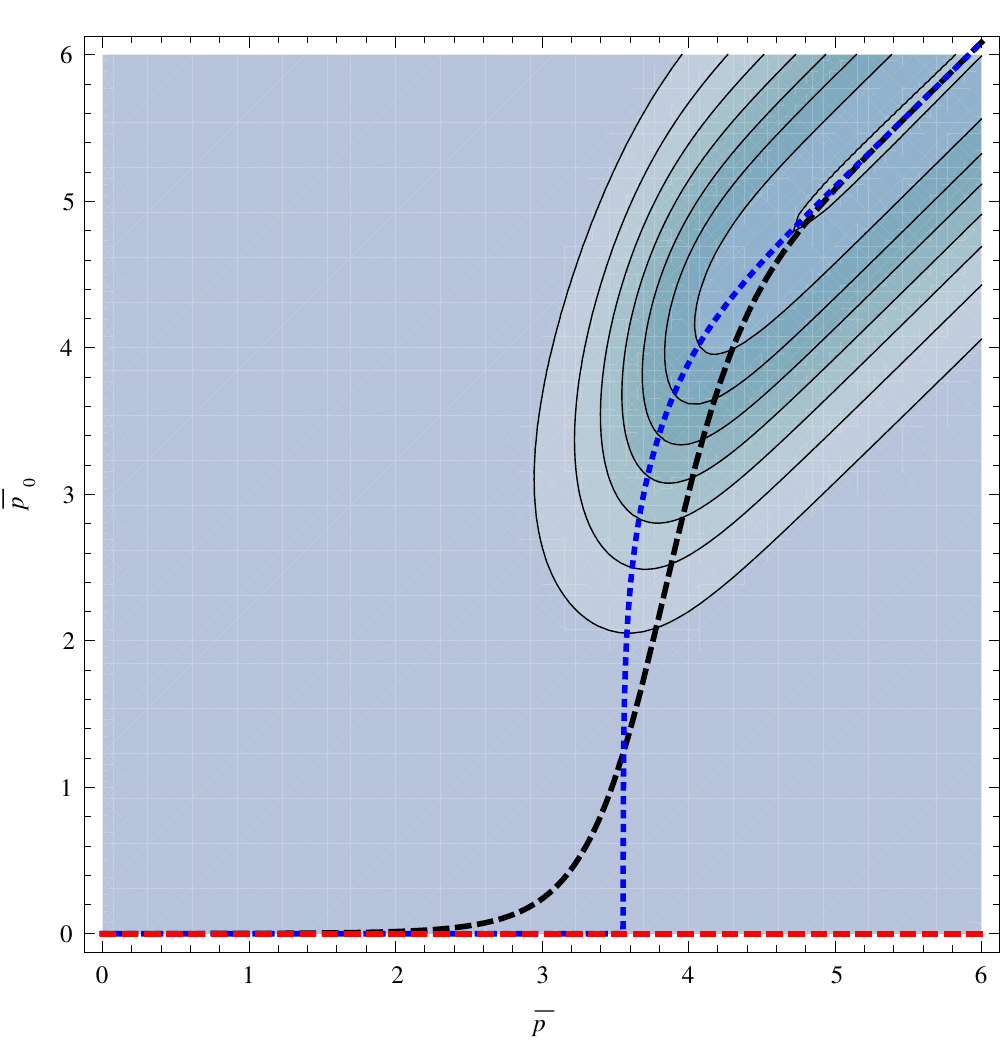}
}
\hspace{0.25cm}
\subfigure
{
    \label{fourpaneljeffd}
    \includegraphics[width=4.1cm]{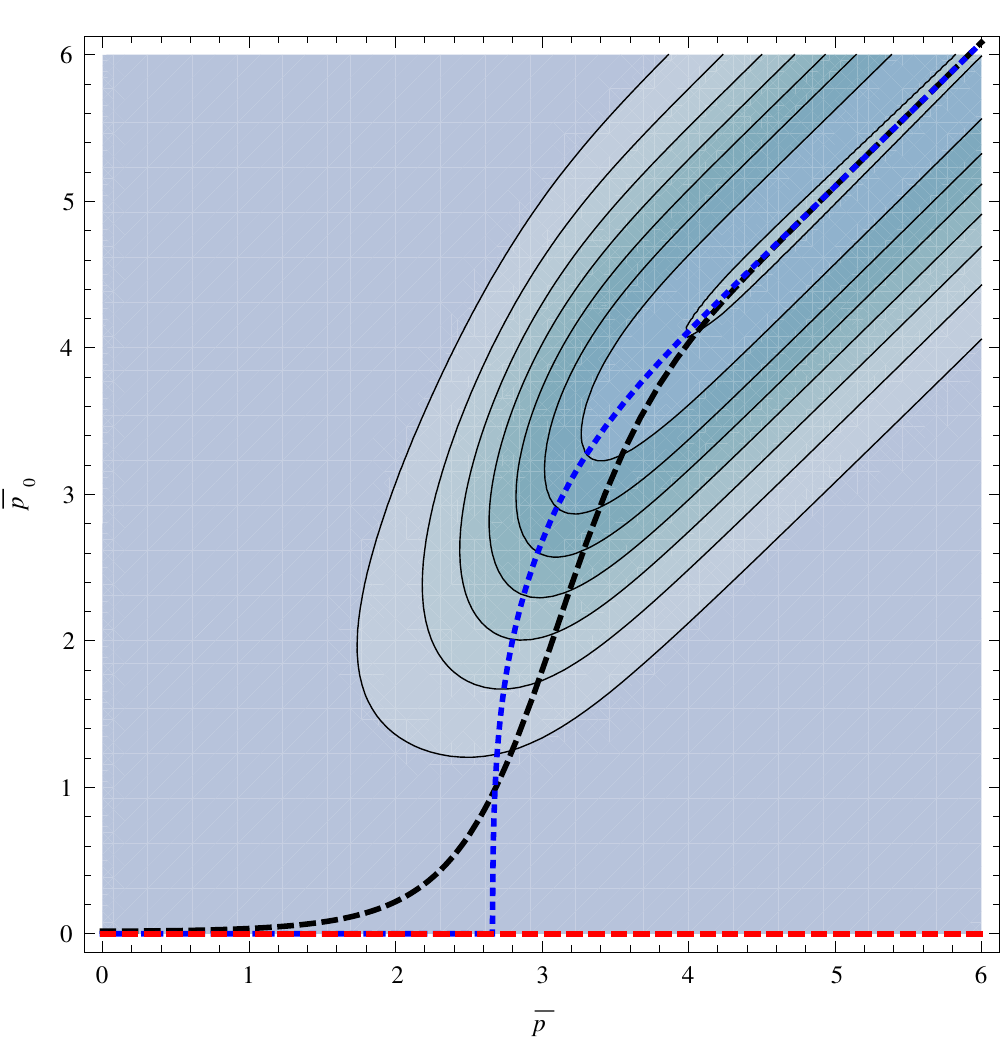}
}
\addtocounter{subfigure}{-1}
\addtocounter{subfigure}{-1}
\hspace{0.25cm}
\subfigure[$A=0.1$ and $\sigma{}=0.01$]
{
    \label{fourpaneljeffcunder}
    \includegraphics[width=4.1cm]{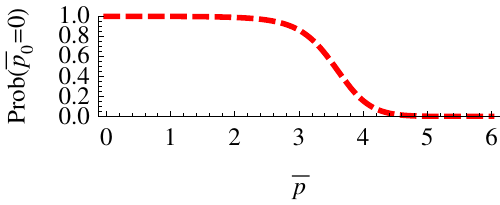}
}
\hspace{0.25cm}
\subfigure[$A=0.1$ and $\sigma{}=0.04$]
{
    \label{fourpaneljeffdunder}
    \includegraphics[width=4.1cm]{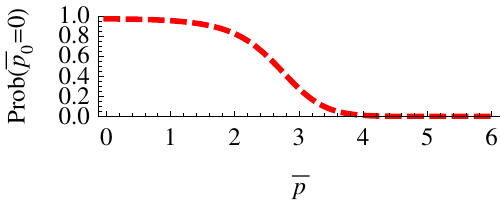}
}
\caption{These four panels show the posterior
distribution $\overline{\rho{}}(\overline{p}_0|\overline{p},\sigma)$ with
prior $\overline{\tau}{}(\overline{p}_0)=2 A \delta(\overline{p}_0) + (1-A)\sigma{}^2 2 \overline{p}_0$
for four combinations of value of $A$ and $\sigma$. The blue dotted line
is the median estimator. The black dashed line is the mean estimator. The red, dashed line at
$\overline{p}_0=0$ is a reminder of the finite amount of probability that
exists along the $\overline{p}$-axis due to the delta function term. No such line is needed in
Figs.~\ref{fig:jeff_contour_rho} and~\ref{fig:jeff_contour_rho_finite}
because $A=0$ there and the delta term does not contribute.
This probability along the $\overline{p}$-axis is plotted underneath each contour plot.}
\label{fourpaneljeff}
\end{figure}

\begin{figure}
\centering
\subfigure
{
    \label{whatevera}
    \includegraphics[width=4.1cm]{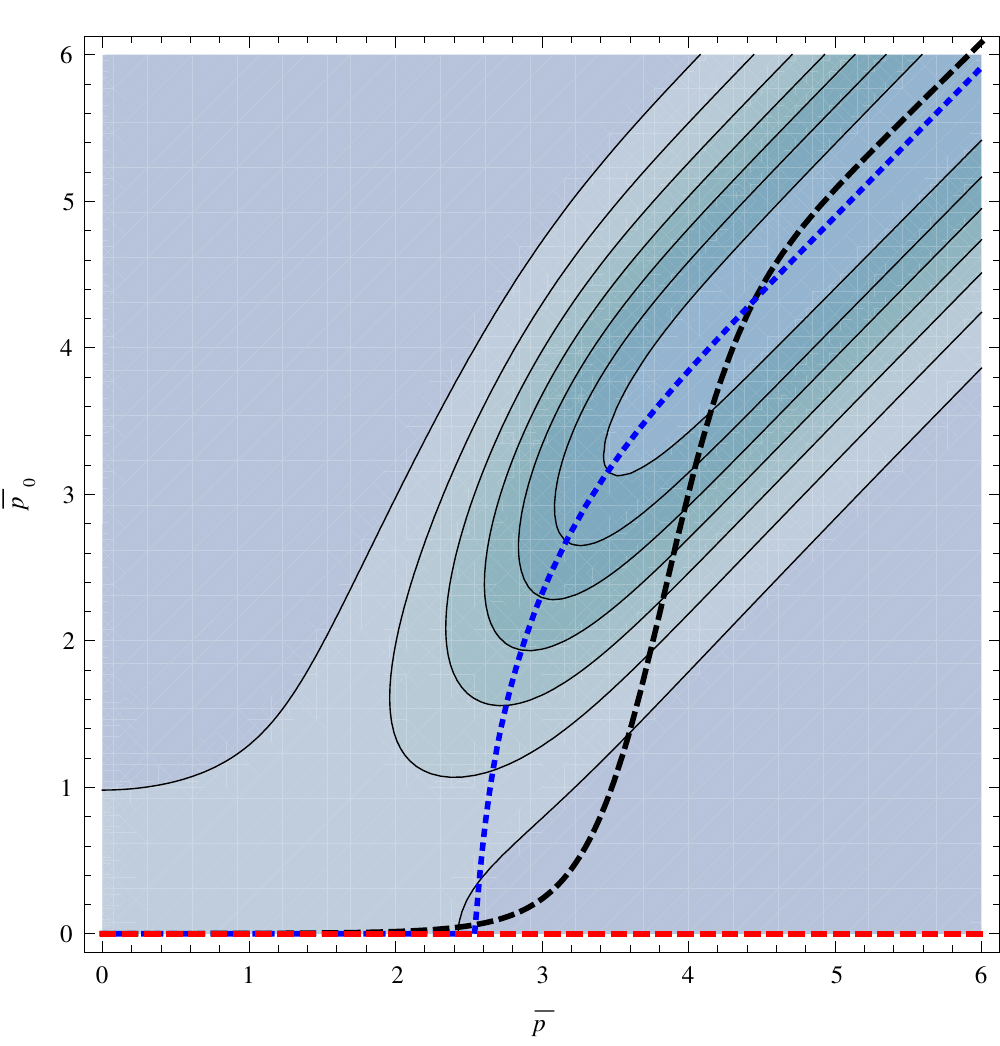}
}
\hspace{0.25cm}
\subfigure
{
    \label{whateverb}
    \includegraphics[width=4.1cm]{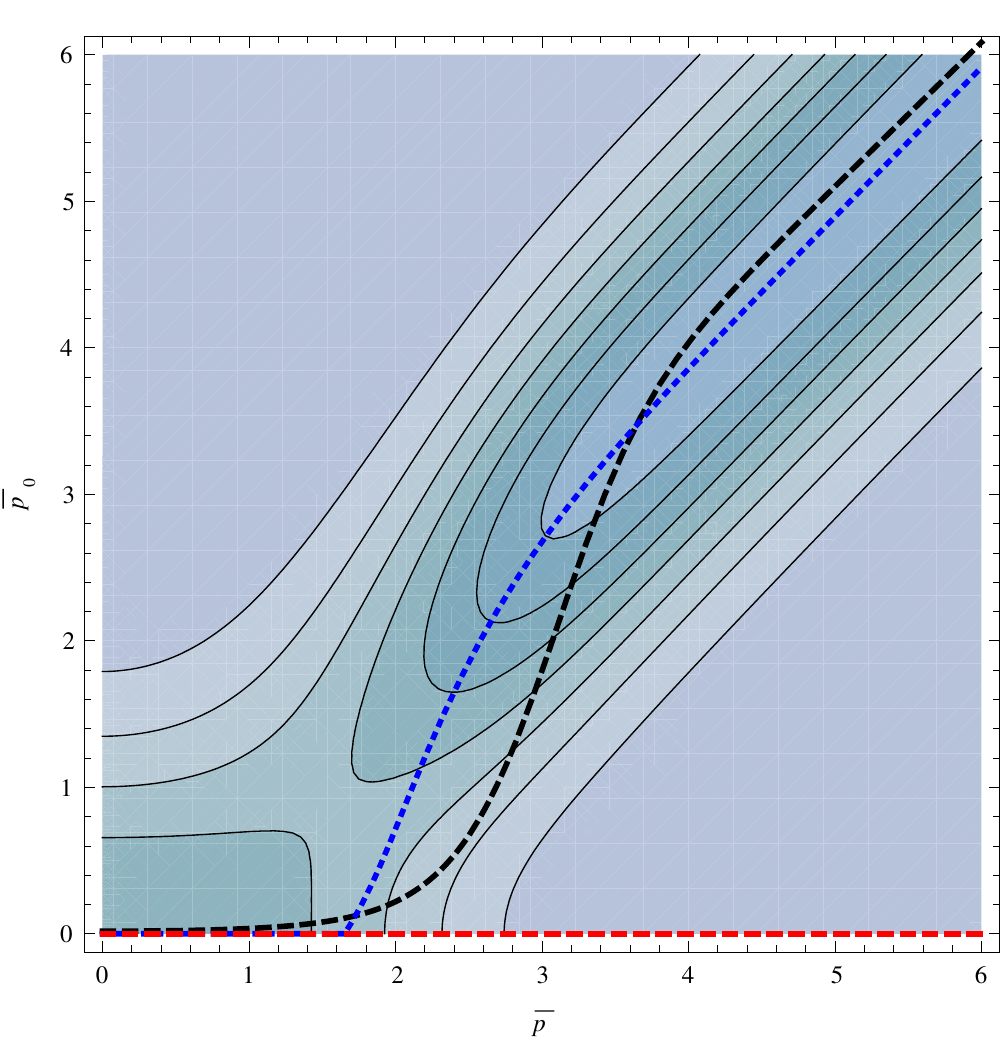}
}
\addtocounter{subfigure}{-1}
\addtocounter{subfigure}{-1}
\hspace{0.25cm}
\subfigure[$A=0.1$ and $\sigma{}=0.01$]
{
    \label{whateveraunder}
    \includegraphics[width=4.1cm]{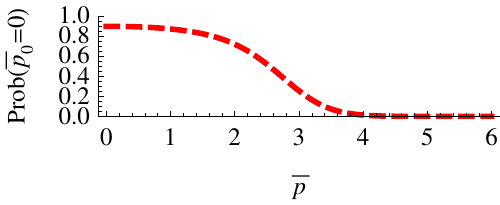}
}
\hspace{0.25cm}
\subfigure[$A=0.1$ and $\sigma{}=0.04$]
{
    \label{whateverbunder}
    \includegraphics[width=4.1cm]{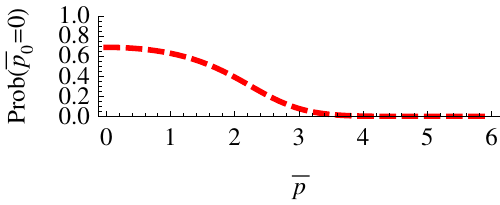}
}
\hspace{0.25cm}
\subfigure
{
    \label{whateverc}
    \includegraphics[width=4.1cm]{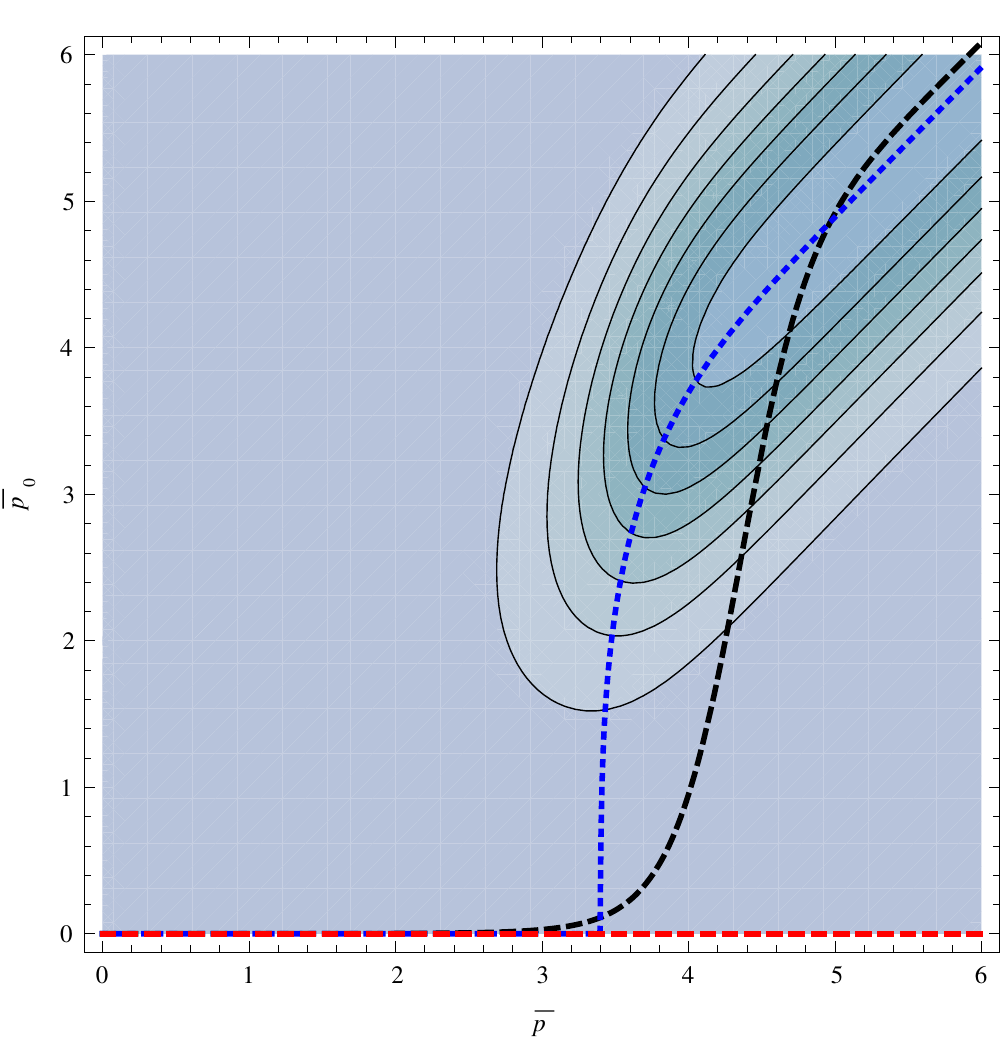}
}
\hspace{0.25cm}
\subfigure
{
    \label{whateverd}
    \includegraphics[width=4.1cm]{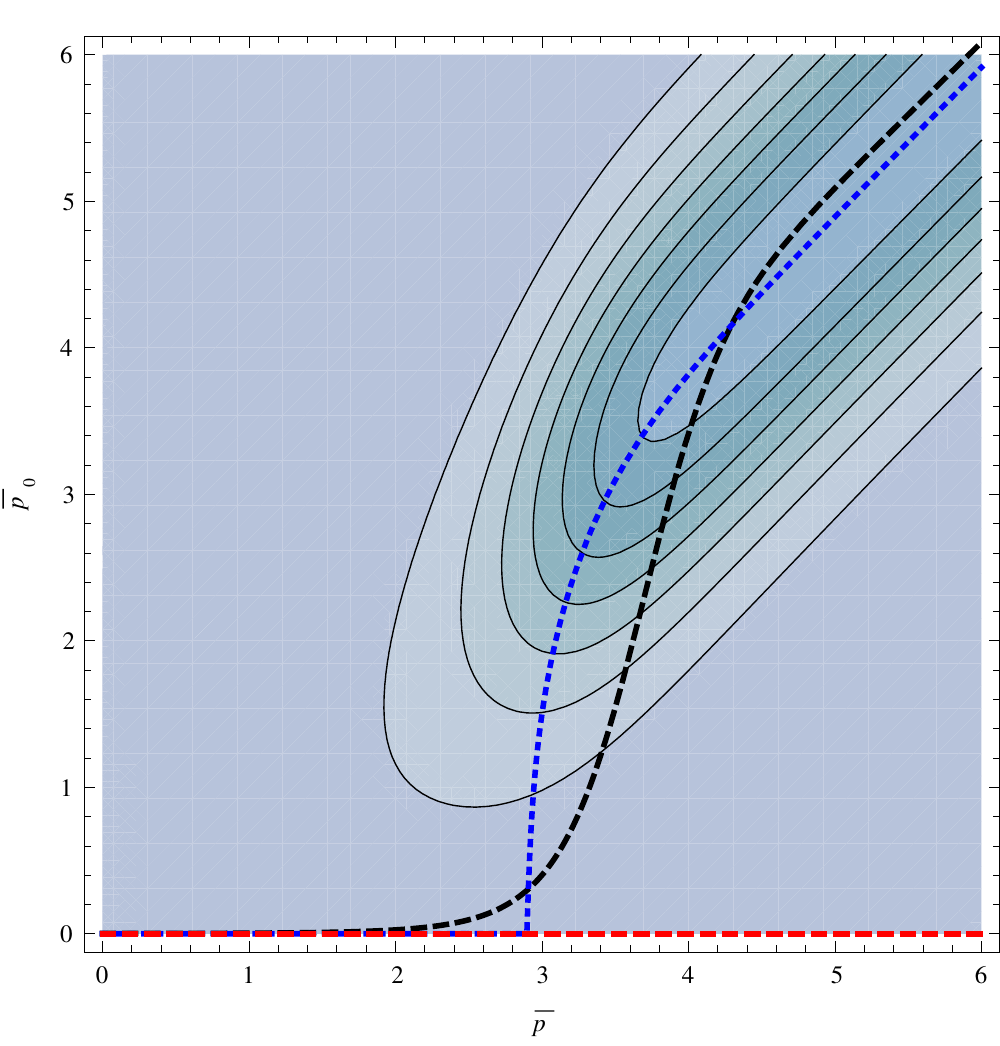}
}
\addtocounter{subfigure}{-1}
\addtocounter{subfigure}{-1}
\hspace{0.25cm}
\subfigure[$A=0.5$ and $\sigma{}=0.01$]
{
    \label{whatevercunder}
    \includegraphics[width=4.1cm]{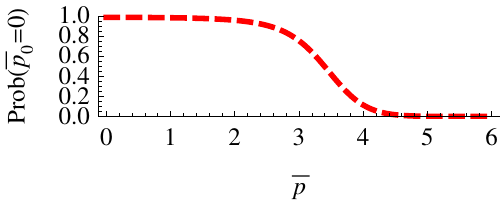}
}
\hspace{0.25cm}
\subfigure[$A=0.5$ and $\sigma{}=0.04$]
{
    \label{whateverdunder}
    \includegraphics[width=4.1cm]{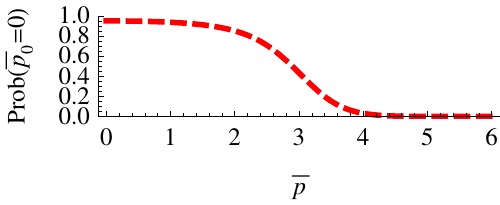}
}
\caption{These four panels show the posterior
distribution $\overline{\rho{}}(\overline{p}_0|\overline{p},\sigma)$ with
prior $\overline{\tau}{}(\overline{p}_0)=2 A \delta(\overline{p}_0) + (1-A)\sigma{}$
for four combinations of value of $A$ and $\sigma$. The blue dotted line
is the median estimator. The black dashed line is the mean estimator. The red, dashed line at
$\overline{p}_0=0$ is a reminder of the finite amount of probability that
exists along the $\overline{p}$-axis due to the delta function term. No such line is needed in
Figs.~\ref{fig:contour_rho} and~\ref{fig:contour_rho_finite}
because $A=0$ there and the delta term does not contribute.
This probability along the $\overline{p}$-axis is plotted underneath each contour plot.}
\label{whatever}
\end{figure}

Figs.~\ref{fourpaneljeff} and~\ref{whatever} show Eq.~\ref{rho_very_general} for the
Jeffreys and uniform priors, respectively, and four different combination of values of $A$ and $\sigma{}$ each.
The red, dashed lines at $\overline{p}_0=0$ are a reminder that Eq.~\ref{rho_very_general}
contains a delta term that cannot be plotted in the usual fashion. Finite probability is attached to
each point along the red line and is graphed beneath each contour plot. This is critical to fully understand these plots.
Several new features are seen for non-zero $A$.
In Fig.~\ref{fourpaneljeff} for the Jeffreys prior, the probability band seems to bend
towards the $\overline{p}$-axis (Fig.~\ref{fourpaneljeffa}) but in actuality abruptly turns and heads towards
the $\overline{p}_0$-axis (Fig.~\ref{fourpaneljeffb}). This
behavior is however strongly damped by residual probability that the polarization is
actually zero (as in Figs.~\ref{fourpaneljeffc} and \ref{fourpaneljeffd}). Fig.~\ref{whatever}
for the uniform polar prior is similar except this time the probability band,
after heading towards the $\overline{p}$-axis as well, turns towards the origin if the
behavior is not damped by the probability that the polarization is zero (as in Figs.~\ref{whateverc} and \ref{whateverd}).

The mean and median estimator curves in Figs.~\ref{fourpaneljeff} and~\ref{whatever} have
changed with the introduction of a non-zero $A$. The mean curve is ``pulled'' more towards
the $\overline{p}$-axis at small values of $\overline{p}$ for larger values of $A$ or smaller
values of $\sigma$. The median curve has a totally new feature: it intercepts the $\overline{p}$-axis
at a critical value, $\overline{p}_{crit}$, and is zero for smaller values.
This occurs when the right-hand side of Eq.~\ref{onedmediansimp} equals zero. Tables of some
critical values, including those corresponding to the figures, are presented in Appendix~\ref{crit}.

When $A=0$ and $\sigma{}=0.01$, a contour plot of Eq.~\ref{rho_very_general} for the Jeffrey
prior reproduces Fig.~\ref{fig:jeff_contour_rho}. This
plot resembles the Rice distribution in Fig.~\ref{fig:contour_rice}, only transposed!
Fig.~\ref{fourpaneljeffb} is a good example of the transitional form that occurs
in the contour shape between the standard $A=0$ form and the form seen
for larger values of $A$ and/or smaller values of $\sigma{}$. This is an important result. It means
that the Rice distribution results can effectively be transposed to find best
estimates of $\overline{p}_0$ for small values of polarization when $1/\sigma \rightarrow \infty$.

When $A=0$ and $\sigma{}=0.01$, a contour plot of Eq.~\ref{rho_very_general} for the uniform
prior reproduces Fig.~\ref{fig:contour_rho}.
There is no delta function and the global maximum of this distribution occurs at the origin.
The wrench-like shape characteristic of this plot arises from small values
of $A$ for a given value of $\sigma$. Fig.~\ref{whateverb} is a good example of the transitional form.

All the usual mathematical machinery may now be applied to $\overline{\rho{}}(\overline{p}_0|\overline{p},\sigma)$
to produce estimators for $\overline{p}_0$ given a measured value of $\overline{p}$.

Previously ``bias corrections'' were performed on measurements at
low signal-to-noise to make them closer to zero. As is seen in the figures with $A=0$,
the mean and median estimators seem to act for very small values of $p$
like a ``bias correction'' that goes the wrong way! (The median estimator
curves with $p$-axis intercepts are an exception.)
Bias corrections should not be used. A non-zero value
for the estimate of $p_0$ even when $p=0$ is not a deficiency with the estimators.
It is a result that can be understood intuitively when all possible values of $p_0$ and $\phi_0$
that could have produced a given $p$ and $\phi$ are considered.

\begin{figure}[t]
\centering
\resizebox{\hsize}{!}{\includegraphics[width=\textwidth]{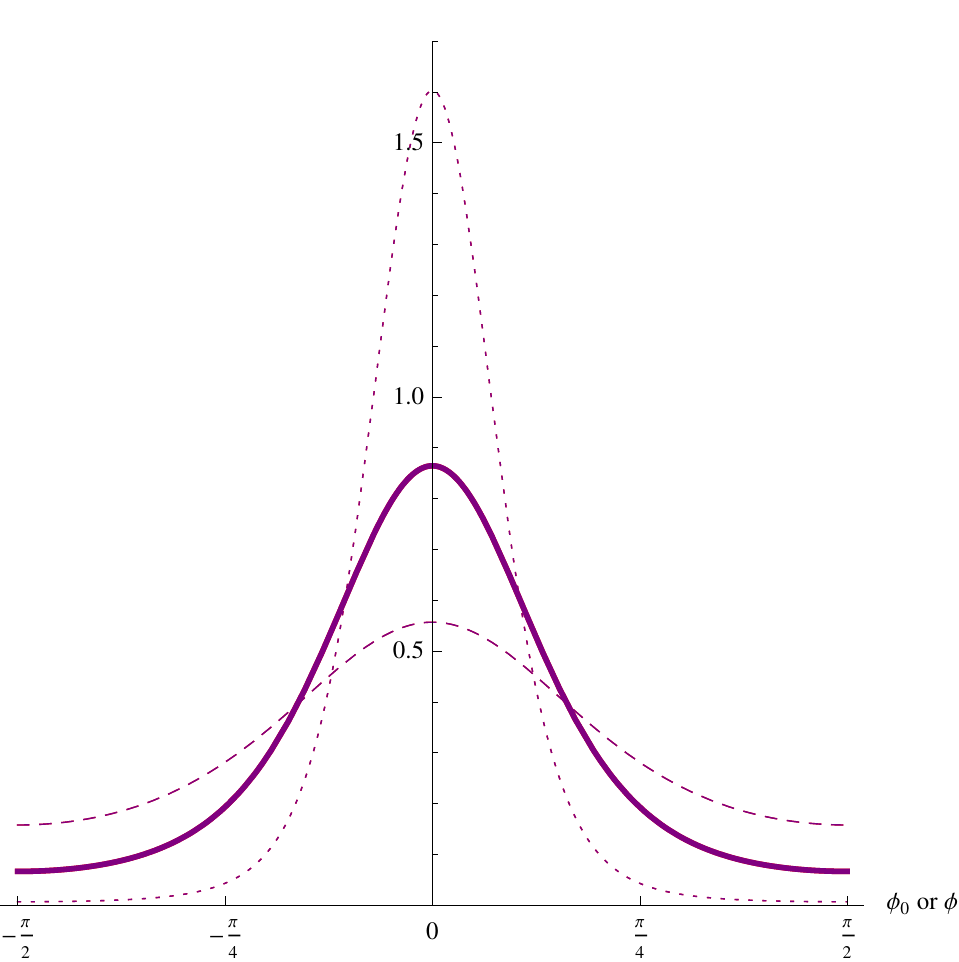}}
\caption{This plot compares the classical angle distribution, $G(\phi|p_0,\phi_0,\sigma)$, to the Bayesian
distribution, $T(\phi_0|p,\phi,\sigma)$, with $\kappa(p_0,\phi{}_0)= 2 p_0/\pi$ and $I_0/\Sigma{}=100$.
The Bayesian curves are virtually identical to the classical curves and overlap in the figure.
Both plots simply use zero for the angle since the choice is immaterial. Three different
values for the signal-to-noise, i.e., $p_0/\sigma{}$ ($G$) or $p/\sigma{}$ ($T$), are used:
$0.5$ (dashed lines), $1.0$ (solid lines), and $2.0$ (dotted lines).}
\label{fig:TvsGjeff}
\end{figure}

\begin{figure}[t]
\centering
\resizebox{\hsize}{!}{\includegraphics[width=\textwidth]{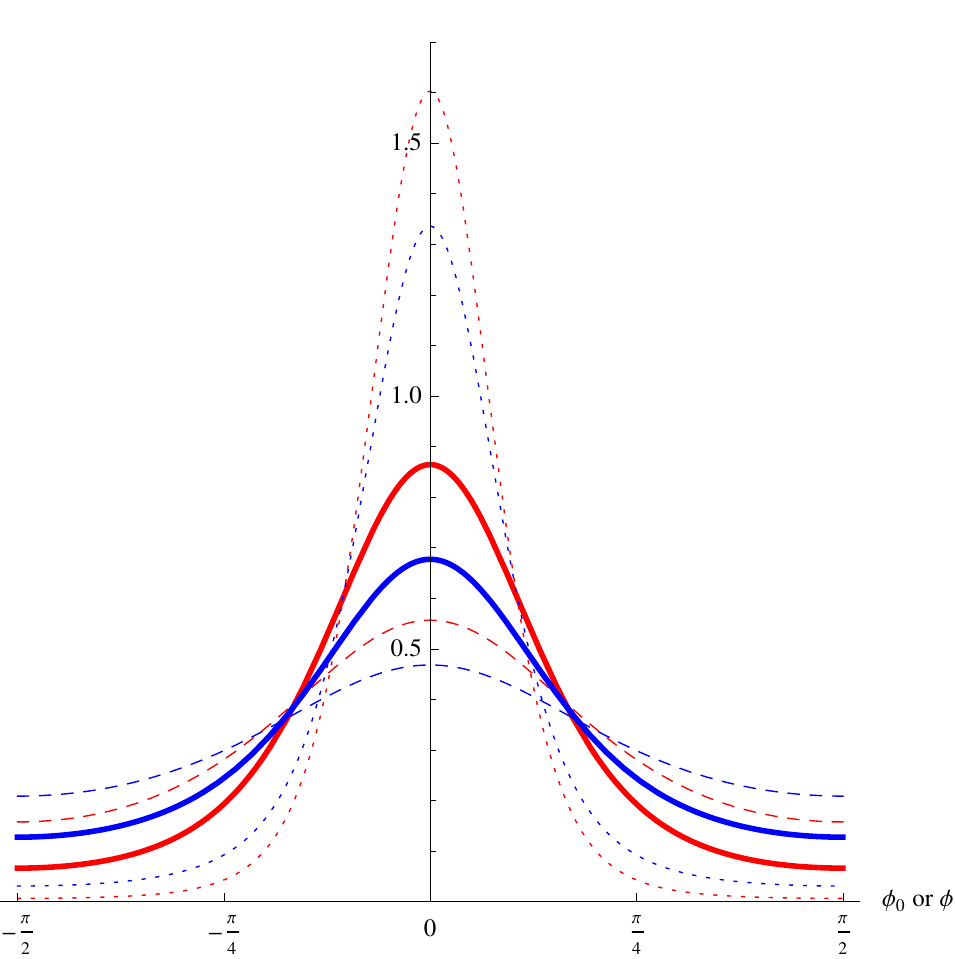}}
\caption{This plot compares the classical angle distribution, $G(\phi|p_0,\phi_0,\sigma)$ (red), to the Bayesian
distribution, $T(\phi_0|p,\phi,\sigma)$ (blue), with $\kappa(p_0,\phi{}_0)=1/\pi$ and $I_0/\Sigma{}=100$.
The Bayesian approach now suggests the classical approach underestimates the error bars. Both
plots simply use zero for the angle since the choice is immaterial. Three different
values for the signal-to-noise, i.e., $p_0/\sigma{}$ ($G$) or $p/\sigma{}$ ($T$), are used:
$0.5$ (dashed lines), $1.0$ (solid lines), and $2.0$ (dotted lines).}
\label{fig:TvsG}
\end{figure}

\section{Marginal distributions of polarization angle}
The marginal posterior distribution, $T$, of $\phi_0$ is given by
\begin{equation}
T(\phi{}_0|p,\phi{},\sigma) = \int_{0}^{1} \! B(p_0,\phi{}_0|p,\phi{},\sigma)  \,   dp_0.
\label{eq:T}
\end{equation}
As before, the denominator of $B$ is a function of $p$ and $\phi{}$ only and 
may be pulled through the integral. The numerator becomes
$\int_0^1 \! f(p,\phi{}|p_0,\phi{}_0,\sigma{}) \kappa{}(p_0,\phi{}_0) \, dp_0$.
If the prior is a function of $p_0$ to some non-negative integer power $n$ (i.e.,
$\kappa{}(p_0) \propto p_0^n$, there appears to exist a family of solutions
via repeated integration by parts. Explicit solutions were found up to $n=10$ using the computer algebra
system Mathematica. Beyond $n=1$, these solutions quickly become impractically large.

The marginal distribution, $G$, of $\phi{}$ over $f$ has been investigated
by several authors (\cite{1965AnAp...28..412V,1986VA.....29...27C,1993A&A...274..968N}).
Its defining formula in terms of the sky angle is
\begin{equation*}
G(\phi{}|p_0,\phi{}_0,\sigma{}) = \int_{0}^\infty{} \! f(p,\phi{}|p_0,\phi{}_0,\sigma{}) \, dp.
\end{equation*}
Unfortunately, the solution to this equation in \cite{1986VA.....29...27C} has some
sign mistakes and the derivation in Appendix~B of \citet{1993A&A...274..968N} confuses
the sky angle, $\phi{}$, with the angle in the $q$-$u$ plane, $\theta{}$.
It is worth explicitly stating both versions to avoid confusion:
\begin{equation}
G(\phi{}|p_0,\phi{}_0,\sigma{}) = \frac{1}{\sqrt{\pi{}}} \left(\frac{1}{\sqrt{\pi{}}} + \eta{}_0 e^{\eta{}^2_0} [1+\operatorname{erf}(\eta{}_0)] \right) e^{-\frac{p^2_0}{2\sigma{}^2}}
\label{eq:G}
\end{equation}
and
\begin{equation}
G'(\theta{}|p_0,\theta{}_0,\sigma{}) = \frac{1}{2\sqrt{\pi{}}} \left(\frac{1}{\sqrt{\pi{}}} + \eta{}'_0 e^{\eta{}'^2_0} [1+\operatorname{erf}(\eta{}'_0)] \right) e^{-\frac{p^2_0}{2\sigma{}^2}},
\end{equation}
where $\eta{}_0=\frac{p_0}{\sqrt{2}\sigma{}} \operatorname{cos}(2(\phi-\phi{}_0))$
and $\eta{}'_0=\frac{p_0}{\sqrt{2}\sigma{}} \operatorname{cos}(\theta-\theta{}_0)$.
It is useful to notice that the distribution is symmetric about $\phi_0$
and using a value of $\phi_0$ different than zero only translates the distribution
along the angular axis. Therefore when calculating confidence intervals, only $\phi_0=0$
need be used.
Unlike the Rice distribution, which is independent of the angular variable, this
distribution depends on $p_0$. $G$ is normalized (i.e., $\int_{-\pi/2}^{\pi{}/2} \! G \, d\phi{} = 1$).

The distribution $T$ cannot in general be related to $G$
as $S$ was to $R$ through $\rho$, even if the prior
is independent of angle, because
$f$ lacks the symmetry in $p$ and $p_0$ that it has for $\phi$ and $\phi_0$.
$T$ and $G$ may have different form. If however $\kappa{}(p_0,\phi_0) \propto p_0$,
as with the Jeffreys prior, it is easy to show that $T \rightarrow G$ as $1/\sigma \rightarrow \infty$.

Fig.~\ref{fig:TvsGjeff}
compares $T$ to $G$ for the Jeffreys prior, i.e., $\kappa{}(p_0,\phi{}_0)= 2 p_0/\pi$,
and $I_0/\Sigma=100$ and three different signal-to-noise values.
As expected, the $T$ distribution and the $G$ distribution are almost identical because $1/\sigma$
is fairly large. The two sets of curves overlap in the figure but the $T$ curves
are not precisely identical to the $G$ curves. The difference between the $T$ and $G$ curves is
larger if $1/\sigma$ is small or for polarization measurements near $1/\sigma$.
In practice, the classical error bars for the angle based on $G$ are perfectly acceptable under this prior.

Fig.~\ref{fig:TvsG} compares $T$ to $G$ for
the case of a uniform polar prior, i.e., $\kappa{}(p_0,\phi{}_0)=1/\pi$, and $I_0/\Sigma=100$.
In the figure,  $T$ (blue) is plotted for $\phi{}=0$ and
$p/\sigma{}=0.5$ (dotted), $1.0$ (solid), $2.0$ (dashed) and
$G$ (red) is plotted for $\phi{}_0=0$ and $p_0/\sigma{}=0.5$ (dotted),
$1.0$ (solid), $2.0$ (dashed). The figure clearly demonstrates that
using $G$, the non-Bayesian formula, to compute errors bars results
in errors bars that are too small. This is especially important
when the polarization degree to $\sigma$ ratio is about one.

If a delta term is added to the priors as was done with the
polarization magnitude plots, the result can roughly be
described by saying that the distribution is scaled by $(1-A)$
and it is then shifted upwards by a constant of $A/\pi$. This
constant shift occurs because the origin has an indeterminate
angle. The equation in barred variables is
\begin{equation}
\overline{T}(\phi{}_0|\overline{p},\phi{},\sigma) =
\frac{\frac{A}{\pi} \frac{\overline{p}}{\pi} \, \operatorname{exp} \left(-\frac{\overline{p}^2}{2}  \right)  + (1-A) \int_0^{1/\sigma} \! \overline{f} \overline{\kappa}_2 \, d\overline{p}_0 }
{A \frac{\overline{p}}{\pi} \, \operatorname{exp} \left(-\frac{\overline{p}^2}{2}  \right) + (1-A) \int_{-\frac{\pi}{2}}^{\frac{\pi}{2}} \int_0^{1/\sigma} \! \overline{f} \overline{\kappa}_2 \, d\overline{p}_0 d\phi_0}.
\end{equation}
When $A=1.0$, the distribution is flat with value $1/\pi$, as it must be
to be normalized.

\section{Summary of derivations}
\begin{figure}
\begin{centering}
\resizebox{\hsize}{!}{\includegraphics[width=\textwidth]{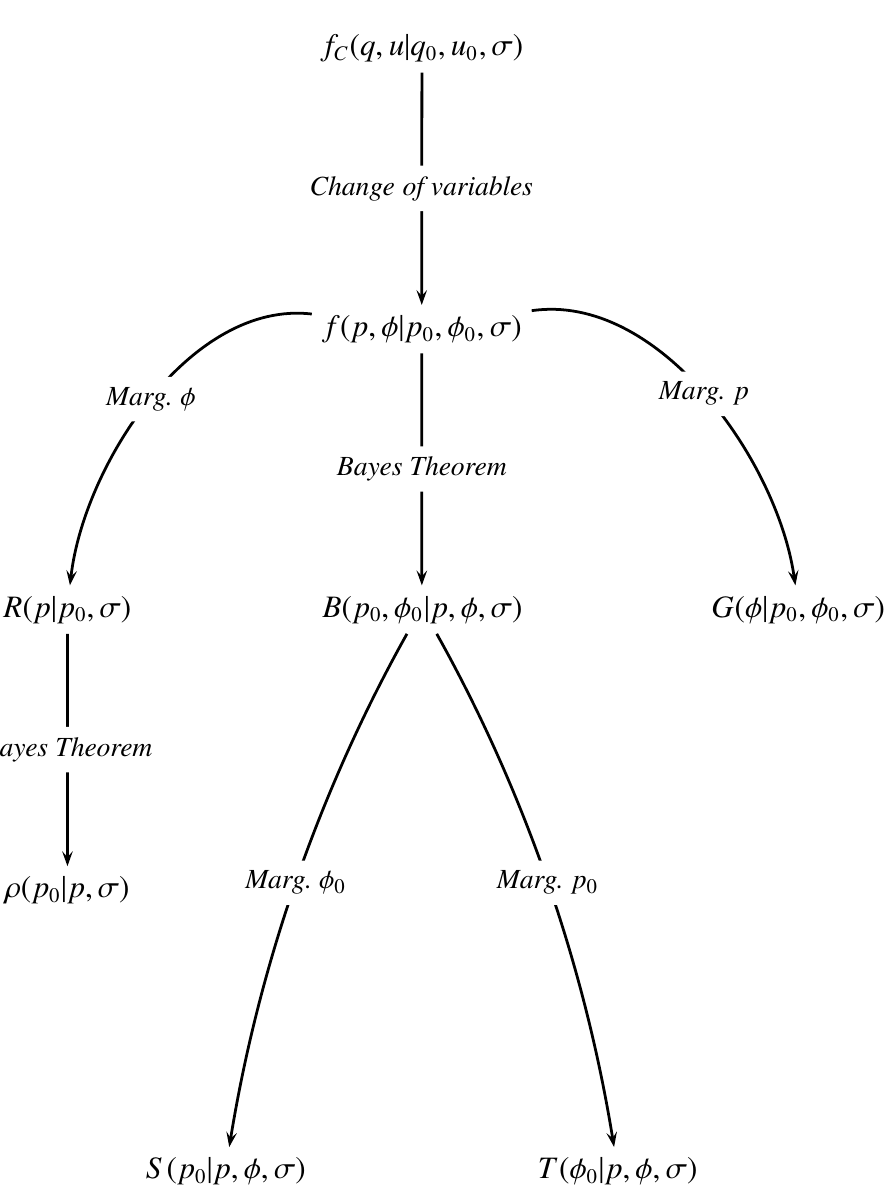}}
\end{centering}
\caption{This figure is a flow chart to help remember the relationships between the distributions presented. ``\textit{Marg.}'' stands for marginalization over the accompanying variable.}
\label{flowchart}
\end{figure}

Fig.~\ref{flowchart} is a flow chart to help remember the relationships
among the many distributions presented. All the main functions result
from a change of variables, Bayes Theorem, or marginalization (denoted
by ``\textit{Marg.}'' in the figure). Attempts were made to remain consistent
with the function names used in previous published papers when possible.

\section{Confidence intervals (aka credible sets)}
All the pieces needed to construct confidence intervals (known
as ``credible sets'' in Bayesian statistics) for $p_0$ and $\phi{}_0$
have been given. In general, these sets will be two-dimensional.
Usually they will be constructed so as to
contain the ``best'' estimate ($\hat{p}_a$ and $\hat{\phi}_a$) that
was found. In practice, especially with polarimetric spectra,
using two-dimensional sets as ``error bars'' is impossible. In these cases,
the marginal distributions, $T$ and $S$, may be used to
construct error bars.

High-quality confidence interval tables, suitable for
use in data reduction pipelines, will be the subject of a future
paper.

\section{Conclusion}
A foundation for a Bayesian treatment of polarization
measurements has been presented based when $V=0$. The treatment covers the
regime where the measured polarization has large or small
Stokes parameters but all the intensities that were used
in the calculation of the Stokes parameters are large enough
that Poisson statistics is unimportant. This justifies the
use of a gaussian approach for the Stokes parameters' error.

A Bayesian analysis of polarization allows the observer
some freedom in choosing a prior and loss function. These
should be stated and justified when presenting one's results. In particular,
the importance of choosing an appropriate prior has been stressed.

The transposed Rice distribution is intimately connected with
the posterior distribution with a two-dimensional Jeffreys prior.
As $1/\sigma \rightarrow \infty$, the distributions are the same.
The classical angular distribution is also the same as the posterior
angular distribution in this limit. A uniform polar prior may be better
for sources with small expected polarization. In this case, the posterior
distributions are rather different than the classical ones.

Further advances may be made by generalizing the analysis to include ellipsoidal,
perhaps even correlated, distributions for $Q$ and $U$, Poisson statistics,
circular polarization, or repeated measurements. Monte Carlo
methods will likely be necessary to handle the integrations for more general
approaches. Polarimetric spectra and time series data that involve measurements
that are not independent and may be highly correlated; modeling
the wavelength-dependence or time-evolution could further improve results
in those situations.

The results presented have focused on the interpretation
of the quantities $p_0$ and $\phi_0$ as polarization
degree and angle but the formulas are applicable to any
application that uses a bounded radial function of the form $\sqrt{x^2+y^2}$,
where $x$ and $y$ are measured with equal gaussian error.

\appendix
\section{Critical values for the median estimator curves with the uniform polar prior}
\label{crit}

Tables~\ref{tableAvsPjeff} and~\ref{tableAvsP} show some critical values
of the median estimator curve for the Jeffreys and uniform polar
prior and an array of choices for $A$ and $1/\sigma$
($=I_0/\Sigma$). Entries with ``$\ldots$'' never
intercept the $\overline{p}$-axis. As is expected,
for a given value of $A$, the critical value
decreases as the data quality increases (i.e.,
$1/\sigma$ gets larger). Also as expected, the
critical values increase as $A$ increases for
a given value of data quality. These tables are created
using Eq.~\ref{onedmediansimp}. The critical values
occur when the right-hand side of that equation is equal to zero.

Unbarred variables are used in the tables.
The behavior in barred variables is somewhat
counter-intuitive and should be avoided to prevent confusion.
As can be seen by comparing the median estimator intercepts for $\overline{p}_0$
in Figs.~\ref{fourpaneljeffc} and~\ref{fourpaneljeffd} and Figs.~\ref{whateverc} and \ref{whateverd},
the intercept is larger for $\sigma=0.01$ than $\sigma=0.04$ even
though the data quality is better for $\sigma=0.01$. This does
not mean that, as the data gets better, a larger measured
value of polarization is required to prove polarization is non-zero!
There is a human tendency to interpret the barred plots as if $\sigma=1$,
but in practice $\sigma$ is always less than $1$. When converted to
unbarred variables, the polarization behaves as expected.

More tables such as this are intended to be included in a future work.

\begin{table}[b]
\caption{Median estimator critical values of $p$ such that $\text{Prob}(p_0=0)=1/2$
for the Jeffreys prior.
}
\label{tableAvsPjeff}
\centering
\begin{tabular}{ l | l l l  l }
\hline\hline
  A  & $I_0/\Sigma=25$ & $I_0/\Sigma=100$ & $I_0/\Sigma=500$ & $I_0/\Sigma=1000$ \\
\hline
 0.01 & 0.060649 & 0.028007 & 0.0075581 & 0.0041300 \\
 0.02 & 0.076999 & 0.030415 & 0.0079219 & 0.0042963 \\
 0.1  & 0.10654  & 0.035553 & 0.0087356 & 0.0046744 \\
 0.2  & 0.11810  & 0.037765 & 0.0090994 & 0.0048450 \\
 0.3  & 0.12519  & 0.039166 & 0.0093331 & 0.0049550 \\
 0.4  & 0.13071  & 0.040278 & 0.0095206 & 0.0050431 \\
 0.5  & 0.13558  & 0.041273 & 0.0096897 & 0.0051231 \\
 0.6  & 0.14029  & 0.042244 & 0.0098556 & 0.0052013 \\
 0.7  & 0.14524  & 0.043277 & 0.010033  & 0.0052856 \\
 0.8  & 0.15106  & 0.044505 & 0.010246  & 0.0053869 \\
 0.9  & 0.15942  & 0.046291 & 0.010558  & 0.0055350 \\
 0.99 & 0.18190  & 0.051210 & 0.011430  & 0.0059525 \\
\hline
\end{tabular}
\end{table}

\begin{table}[b]
\caption{Median estimator critical values of $p$ such that $\text{Prob}(p_0=0)=1/2$
for the uniform polar prior. Entries marked ``$\ldots$'' mean that $\text{Prob}(p_0=0)<1/2$ for
all values of $p$ so no solution exists for $p_{\text{crit}}$.
}
\label{tableAvsP}
\centering
\begin{tabular}{ l | l l l  l }
\hline\hline
  A  & $I_0/\Sigma=25$ & $I_0/\Sigma=100$ & $I_0/\Sigma=500$ & $I_0/\Sigma=1000$ \\
\hline
 0.01 & $\ldots$ &  $\ldots$ & 0.0042099 & 0.0024939 \\
 0.02 & $\ldots$ &  0.013263  & 0.0049980 & 0.0028143 \\
 0.1 & 0.066096 & 0.025408 & 0.0063962 & 0.0034318 \\
 0.2 & 0.089414 & 0.028977 & 0.0069392 & 0.0036822 \\
 0.3 & 0.10092 & 0.031047 & 0.0072724 & 0.0038378 \\
 0.4 & 0.10910 & 0.032617 & 0.0075324 & 0.0039599 \\
 0.5 & 0.11591 & 0.033976 & 0.0077618 & 0.0040683 \\
 0.6 & 0.12221 & 0.035270 & 0.0079835 & 0.0041734 \\
 0.7 & 0.12861 & 0.036617 & 0.0082172 & 0.0042846 \\
 0.8 & 0.13590 & 0.038183 & 0.0084923 & 0.0044159 \\
 0.9 & 0.14603 & 0.040404 & 0.0088880 & 0.0046056 \\
 0.99 & 0.17197  &  0.046273 & 0.0099570 & 0.0051216 \\
\hline
\end{tabular}
\end{table}

\section{``Projected'' prior}
\label{projappend}
A third interesting prior was discussed in Section~\ref{Elicitation}. This prior, called
here the ``projected'' prior, is motivated by considering an instrument that only measures
linear polarization but ignores circular polarization. While a distribution for $q$ and $u$
can be derived for this case and the machinery given in the main text can be applied to it,
this is somewhat inappropriate since it violates the leading assumption that $V=0$ upon
which the subsequent equations were derived. Nevertheless, it is worth recording here some
of the salient formulas.

\begin{figure}[t]
\centering
\resizebox{\hsize}{!}{\includegraphics[width=\textwidth]{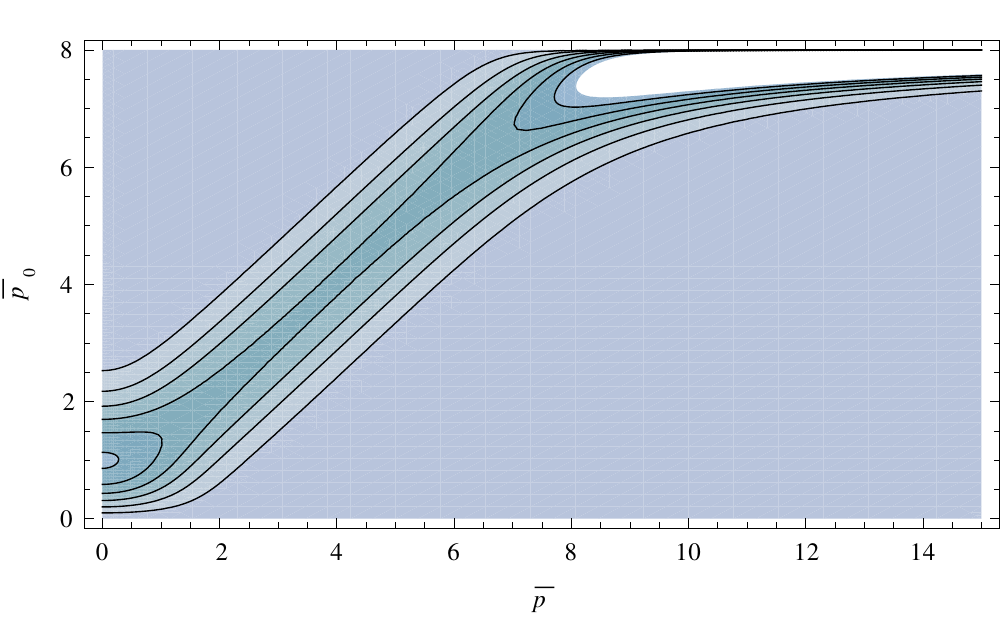}}
\caption{The posterior Rice distribution using the ``projected'' prior
of Eq.~\ref{projectedbarred} with $I_0/\Sigma{}=8$ (a very small value). The
distribution of the measured values near the physically-maximum value is
notably different than the previous corresponding figures for the Jeffreys and uniform
polar prior but the small-value behavior is nearly exact to the Jeffreys case.}
\label{fig:projected}
\end{figure}

The starting point is a normalized uniform Cartesian prior
on the Poincar\'{e} sphere
\begin{equation}
\kappa_C(q_0,u_0,v_0)=\frac{3}{4 \pi{}}.
\end{equation}
Upon projection onto the $q$-$u$ plane, the probability
density function becomes
\begin{equation}
\kappa_{proj}(q_0,u_0)=\frac{3}{2 \pi{}} \sqrt{1-q_0^2-u_0^2}.
\end{equation}
Equivalently in polar coordinates, this is
\begin{equation}
\kappa_{proj}(p_0,\phi_0)=\frac{3}{\pi{}} p_0 \sqrt{1-p_0^2}
\end{equation}
or
\begin{equation}
\overline{\kappa}_{proj}(\overline{p}_0,\phi_0)=\frac{3 \sigma{}^2}{\pi{}} \overline{p}_0 \sqrt{1-\sigma{}^2\overline{p}_0^2}.
\end{equation}
Notice that for small values of $\overline{p}_0$, Eq.~\ref{projectedbarred} has the same form
as Eq.~\ref{polarbarred} up to an irrelevant scale factor. Thus
for small values of $\overline{p}_0$, the posterior distribution should be
similar to that of the Jeffreys posterior. The large value behavior
will be slightly different.
Marginalization of $\overline{\kappa}_{proj}(\overline{p}_0,\phi_0)$ over $\phi$ produces
\begin{equation}
\overline{\tau}_{proj}(\overline{p}_0)=3 \sigma{}^2 \overline{p}_0 \sqrt{1-\sigma{}^2\overline{p}_0^2}
\label{projectedbarred}
\end{equation}
which is plotted in Fig.~\ref{fig:projected} and may be contrasted with
the previous similar figures.

\begin{acknowledgements}
The authors acknowledge support from the Millennium Center for
Supernova Science (MCSS) through grant P10-064-F funded by ``Iniciativa Cient\'ifica Milenio''. The contour
plots were generated using computer algebra system \textit{Mathematica} (version 7.01.0).
The authors wish to thank Alejandro Clocchiatti and Lifan Wang for helpful discussion
and the anonymous referee for valuable comments.
\end{acknowledgements}

\bibliographystyle{aa}
\bibliography{poltheory}

\begin{thebibliography}{25}
\expandafter\ifx\csname natexlab\endcsname\relax\def\natexlab#1{#1}\fi

\bibitem[{{Berger}(1985)}]{1985bergerbook}
{Berger}, J.~O. 1985, {Statistical Decision Theory and Bayesian Analsyis,
  Second Edition} ({Springer})

\bibitem[{{Bertrand}(1888)}]{1889Bertrandbook}
{Bertrand}, J. 1888, {Calcul des probabilit{\'e}s.} ({Gauthier-Villars. Paris})

\bibitem[{{Clarke} \& {Stewart}(1986)}]{1986VA.....29...27C}
{Clarke}, D. \& {Stewart}, B.~G. 1986, Vistas in Astronomy, 29, 27

\bibitem[{{Clarke} {et~al.}(1983){Clarke}, {Stewart}, {Schwarz}, \&
  {Brooks}}]{1983A&A...126..260C}
{Clarke}, D., {Stewart}, B.~G., {Schwarz}, H.~E., \& {Brooks}, A. 1983, \aap,
  126, 260

\bibitem[{{del Toro Iniesta}(2003)}]{2003isp..book.....D}
{del Toro Iniesta}, J.~C. 2003, {Introduction to Spectropolarimetry}
  ({Cambridge University Press})

\bibitem[{{Di Porto} {et~al.}(2010){Di Porto}, {Crosignani}, {Ciattoni}, \&
  {Liu}}]{2010arXiv1008.1878D}
{Di Porto}, P., {Crosignani}, B., {Ciattoni}, A., \& {Liu}, H.~C. 2010, ArXiv
  e-prints

\bibitem[{{Jaynes}(1973)}]{1973FoPh....3..477J}
{Jaynes}, E.~T. 1973, Foundations of Physics, 3, 477

\bibitem[{Jeffreys(1939)}]{Jeffreys1939theory}
Jeffreys, H. 1939, {Theory of probability} (Oxford University Press)

\bibitem[{Jeffreys(1946)}]{1946Jeffries}
Jeffreys, H. 1946, Proceedings of the Royal Society of London. Series A,
  Mathematical and Physical Sciences, 186, pp. 453

\bibitem[{{Keller}(2002)}]{2002apsp.conf..303K}
{Keller}, C.~U. 2002, in Astrophysical Spectropolarimetry, ed.
  {J.~Trujillo-Bueno, F.~Moreno-Insertis, \& F.~S{\'a}nchez}, 303--354

\bibitem[{{Landi Degl'Innocenti}(2002)}]{2002apsp.conf....1L}
{Landi Degl'Innocenti}, E. 2002, in Astrophysical Spectropolarimetry, ed.
  {J.~Trujillo-Bueno, F.~Moreno-Insertis, \& F.~S{\'a}nchez}, 1--53

\bibitem[{{Maronna} {et~al.}(1992){Maronna}, {Feinstein}, \&
  {Clocchiatti}}]{1992A&A...260..525M}
{Maronna}, R., {Feinstein}, C., \& {Clocchiatti}, A. 1992, \aap, 260, 525

\bibitem[{{Naghizadeh-Khouei} \& {Clarke}(1993)}]{1993A&A...274..968N}
{Naghizadeh-Khouei}, J. \& {Clarke}, D. 1993, \aap, 274, 968

\bibitem[{{Patat} \& {Romaniello}(2006)}]{2006PASP..118..146P}
{Patat}, F. \& {Romaniello}, M. 2006, \pasp, 118, 146

\bibitem[{{Rice}(1945)}]{1945BSTJ...24...46R}
{Rice}, S.~O. 1945, Bell Systems Tech.~J., Volume 24, p.~46-156, 24, 46

\bibitem[{{Robert}(1994)}]{1994rogerbook}
{Robert}, C.~P. 1994, {The Bayesian Choice: A Decision-Theoretic Motivation}
  ({Springer})

\bibitem[{{Robert} {et~al.}(2008){Robert}, {Chopin}, \&
  {Rousseau}}]{2008arXiv0804.3173R}
{Robert}, C.~P., {Chopin}, N., \& {Rousseau}, J. 2008, ArXiv e-prints

\bibitem[{{Robert} {et~al.}(2009){Robert}, {Chopin}, \&
  {Rousseau}}]{2009arXiv0909.1008R}
{Robert}, C.~P., {Chopin}, N., \& {Rousseau}, J. 2009, ArXiv e-prints

\bibitem[{{Serkowski}(1958)}]{1958AcA.....8..135S}
{Serkowski}, K. 1958, Acta Astronomica, 8, 135

\bibitem[{{Simmons} \& {Stewart}(1985)}]{1985A&A...142..100S}
{Simmons}, J.~F.~L. \& {Stewart}, B.~G. 1985, \aap, 142, 100

\bibitem[{Tissier(1984)}]{1984Tissier}
Tissier, P.~E. 1984, The Mathematical Gazette, 68, pp. 15

\bibitem[{{Vaillancourt}(2006)}]{2006PASP..118.1340V}
{Vaillancourt}, J.~E. 2006, \pasp, 118, 1340

\bibitem[{{Vinokur}(1965)}]{1965AnAp...28..412V}
{Vinokur}, M. 1965, Annales d'Astrophysique, 28, 412

\bibitem[{{Wang} {et~al.}(1997){Wang}, {Wheeler}, \&
  {Hoeflich}}]{1997ApJ...476L..27W}
{Wang}, L., {Wheeler}, J.~C., \& {Hoeflich}, P. 1997, \apjl, 476, L27+

\bibitem[{{Wardle} \& {Kronberg}(1974)}]{1974ApJ...194..249W}
{Wardle}, J.~F.~C. \& {Kronberg}, P.~P. 1974, \apj, 194, 249

\end{thebibliography}

\end{document}